\documentclass[twocolumn]{aastex63}
\usepackage[utf8]{inputenc}
\usepackage{hyperref}
\usepackage{float}
\usepackage{verbatim}
\usepackage{natbib}
\usepackage[caption=false]{subfig}
\usepackage{multirow}
\usepackage{dcolumn}
\usepackage[nomarkers,nolists]{endfloat}
\newcolumntype{d}[1]{D{.}{.}{#1}}

\usepackage{xcolor}

\newcommand\Nquads{101~}

\begin{document}

\title{\Nquads Eclipsing Quadruple Star Candidates Discovered in {\em TESS} Full Frame Images}

\correspondingauthor{Veselin B. Kostov}
\email{veselin.b.kostov@nasa.gov}
\author[0000-0001-9786-1031]{Veselin~B.~Kostov}
\affiliation{NASA Goddard Space Flight Center, 8800 Greenbelt Road, Greenbelt, MD 20771, USA}
\affiliation{SETI Institute, 189 Bernardo Ave, Suite 200, Mountain View, CA 94043, USA}
%
\author[0000-0003-0501-2636]{Brian P. Powell}
\affiliation{NASA Goddard Space Flight Center, 8800 Greenbelt Road, Greenbelt, MD 20771, USA}
\author[0000-0003-3182-5569]{Saul A. Rappaport}
\affiliation{Department of Physics, Kavli Institute for Astrophysics and Space Research, M.I.T., Cambridge, MA 02139, USA}
\author[0000-0002-8806-496X]{Tam\'as Borkovits}
\affiliation{Baja Astronomical Observatory of University of Szeged, H-6500 Baja, Szegedi út, Kt. 766, Hungary}
\affiliation{HUN-REN -- SZTE Stellar Astrophysics Research Group,  H-6500 Baja, Szegedi út, Kt. 766, Hungary}
\affiliation{Konkoly Observatory, Research Centre for Astronomy and Earth Sciences, H-1121 Budapest, Konkoly Thege Miklós út 15-17, Hungary}
\affiliation{ELTE Gothard Astrophysical Observatory, H-9700 Szombathely, Szent Imre h. u. 112, Hungary}
\author[0000-0002-5665-1879]{Robert Gagliano}
\affiliation{Amateur Astronomer, Glendale, AZ 85308}
\author[0000-0003-3988-3245]{Thomas L. Jacobs}
\affiliation{Amateur Astronomer, 12812 SE 69th Place, Bellevue, WA 98006}
\author[0000-0002-7778-3117]{Rahul Jayaraman}
\affiliation{Department of Physics, Kavli Institute for Astrophysics and Space Research, M.I.T., Cambridge, MA 02139, USA}
\author[0000-0002-2607-138X]{Martti~H.~Kristiansen}
\affil{Brorfelde Observatory, Observator Gyldenkernes Vej 7, DK-4340 T\o{}ll\o{}se, Denmark}
%
\author[0000-0002-8527-2114]{Daryll M. LaCourse}
\affiliation{Amateur Astronomer, 7507 52nd Place NE Marysville, WA 98270}
\author{Tibor Mitnyan}
\affiliation{ELKH–SZTE Stellar Astrophysics Research Group, H-6500 Baja, Szegedi \'ut, Kt. 766, Hungary}
\author{Mark Omohundro}
\affiliation{Citizen Scientist, c/o Zooniverse, Department of Physics, University of Oxford, Denys Wilkinson Building, Keble Road, Oxford, OX13RH, UK}
\author{Jerome Orosz}
\affiliation{Department of Astronomy, San Diego State University, 5500 Campanile Drive, San Diego, CA 92182, USA}
\author{Andr\'as P\'al}
\affiliation{Konkoly Observatory, Research Centre for Astronomy and Earth Sciences, MTA Centre of Excellence, Konkoly Thege Mikl\'os  \'ut 15-17, H-1121 Budapest, Hungary}
\author[0000-0002-5034-0949]{Allan R. Schmitt}
\affiliation{Citizen Scientist, 616 W. 53rd. St., Apt. 101, Minneapolis, MN 55419, USA}
\author{Hans M. Schwengeler}
\affiliation{Citizen Scientist, Planet Hunter, Bottmingen, Switzerland}
\author{Ivan A. Terentev}
\affiliation{Citizen Scientist, Planet Hunter, Petrozavodsk, Russia}
\author[0000-0002-5286-0251]{Guillermo Torres}
\affiliation{Center for Astrophysics $\vert$ Harvard \& Smithsonian, 60 Garden St, Cambridge, MA, 02138, USA}
\author{Thomas Barclay}
\affiliation{NASA Goddard Space Flight Center, 8800 Greenbelt Road, Greenbelt, MD 20771, USA}
\affiliation{University of Maryland, Baltimore County, 1000 Hilltop Circle, Baltimore, MD 21250, USA}
%
%
%
%
\author[0000-0001-7246-5438]{Andrew Vanderburg}
\affiliation{Department of Astronomy, University of Wisconsin-Madison, Madison, WI 53706, USA}
\author{William Welsh}
\affiliation{Department of Astronomy, San Diego State University, 5500 Campanile Drive, San Diego, CA 92182, USA}
\begin{abstract}
We present our second catalog of quadruple star candidates, containing \Nquads systems discovered in {\em TESS} Full-Frame Image data. The targets were initially detected as eclipsing binary stars with the help of supervised machine learning methods applied to sectors Sectors 1 through 54. A dedicated team of citizen scientists subsequently identified through visual inspection two sets of eclipses following two different periods. All \Nquads systems presented here pass comprehensive photocenter motion tests confirming that both sets of eclipses originate from the target star. Some of the systems exhibit prominent eclipse time variations suggesting dynamical interactions between the two component binary stars. One target is an eclipsing quintuple candidate with a (2+1)+2 hierarchical configuration, such that the (2+1) subsystem produces eclipses on the triple orbit as well. Another has recently been confirmed as the second shortest period quadruple reported to date. This catalog provides ephemerides, eclipse depths and durations, sample statistics, and highlights potentially interesting targets for future studies. 
\end{abstract}

\accepted{MNRAS Sep 2023}


\section{Introduction}\label{sec:intro}

Most of the stars more massive than the Sun reside in binary or multiple stellar systems, such that the multiplicity fraction increases with mass \citep[e.g.][]{2010ApJS..190....1R, 2018ApJS..235....6T, 2017ApJS..230...15M}. Eclipsing binaries represent a small fraction of all binary stars yet provide an outsized contribution to stellar astrophysics as fundamental calibrators for stellar masses, sizes, temperatures, and luminosities \citep[e.g.][]{1991A&ARv...3...91A, 2010A&ARv..18...67T}. The contribution is even greater from eclipsing binaries (EBs) found in e.g. 2+1, (2+1)+1, 2+2, (2+2)+2 hierarchical configurations, especially when the systems are compact enough for the various components to interact with each other on observable timescales \citep[e.g.][]{2016MNRAS.455.4136B,2022MNRAS.515.3773B,2022MNRAS.510.1352B,2022Galax..10....9B,2017MNRAS.467.2160R,2023MNRAS.521..558R,2022MNRAS.513.4341R,2021AJ....161..162P,2021ApJ...917...93K,2023MNRAS.522...90K, 2015ASPC..496...55O}. The interactions can be dynamical, evolutionary, or both, and can disrupt the system, trigger common envelope events, result in mergers, and even produce supernovae events \citep[e.g.][]{1962P&SS....9..719L,1962AJ.....67..591K,2013MNRAS.435..943P,2018MNRAS.476.4234F,2021MNRAS.502.4479H,2019MNRAS.486.4781F,2019MNRAS.483.4060L,2022PhRvD.106d3014T,2022ApJ...926..195V,2021MNRAS.507.5832K,2021MNRAS.507..560S}. One may wonder what would the reaction of \cite{1783RSPT...73..474G,1881HarPa...1...15P}, and \cite{1890AN....123..289V} be had they known that Algol, the archetype eclipsing binary they studied in depth, is in fact a triple star.  

Close binary stars provide an excellent discovery opportunity for multiple stellar systems through detected radial velocity and/or eclipse time variations, or even the presence of additional eclipses. Such discoveries bring deeper understanding of, for example, the orbital and physical properties of 2+2 quadruple systems which, in turn, can shed light on their origin by helping to distinguish between gravitational capture or core/disk fragmentation \citep[e.g.][]{1994ARA&A..32..465M,2015Natur.518..213P,2016Natur.538..483T,Tokovinin2021,Whitworth2001,2023ApJS..264...45F,2022MNRAS.517.2111P,2022MNRAS.516.1406S}. For example, a compact quadruple system where the two sub-systems and their component stars are of comparable mass, and the mutual inclination is small may suggest a close encounter between two protobinary stars followed by disk fragmentation \citep[e.g.][]{2021ApJ...917...93K}. 

Comprehensive studies of the formation and evolution of multiple stellar systems depend on systematic surveys exploring a variety of stellar populations and Galactic environments. The last two centuries of EB studies have culminated in large ground- and space-based photometric surveys such as e.g. ASAS-SN, ATLAS, CoRoT, Gaia, Kepler, OGLE, and TESS, that have collectively observed hundreds of thousands of EBs \citep[e.g.][]{2017PASP..129j4502K,2018AJ....156..241H,2022MNRAS.517.2190R,2011AJ....141...83P,2022ApJS..258...16P,2018A&A...619A..97D,2022MNRAS.509..246H,2022arXiv221100929M,2016AcA....66..405S}. Combined with multiple efforts to digitize plate archives, and thus extend the observed baseline as far back as the end of the 19th century \citep[e.g.][]{2012IAUS..285...29G}, this comprehensive treasure trove of data has rejuvenated and revolutionized the field of eclipsing binary stars, and enabled major contributions to our understanding of stars and planets in multiple stellar systems. 

The Transiting Exoplanet Survey Satellite (TESS, \citealt{2015JATIS...1a4003R}), in particular, has been monitoring the sky for transiting exoplanets since 2018, and has already detected thousands of exoplanet candidates (more than 9,700 at the time of writing according to the Exoplanet Follow-up Observing Program archive\footnote{\url{https://exofop.ipac.caltech.edu/tess/}}). As an all-sky photometric survey, it is also ideally-suited to observe a large number of eclipsing binary stars, as well as various other astrophysical variable sources ranging from Solar System objects to Active Galactic Nuclei and beyond. Tens of thousands of TESS EBs have already been discovered through dedicated efforts or as a byproduct of planet searches \citep[e.g.][]{2022ApJS..258...16P,2022RNAAS...6...96H,2023MNRAS.522...29G,2022BAAS...54e.414E,2023MNRAS.518L..31M,2023arXiv230206724M,2022MNRAS.513..102C,2023MNRAS.521.3749M} and many more are expected to come in the future \citep[e.g.][]{2015ApJ...809...77S,2021tsc2.confE.163K}.  

Capitalizing on TESS data we have been searching for EBs, triple, quadruple and higher-order systems, circumbinary planets, and other unusual systems. The search is refined by machine learning, powered by citizen scientists, and based on Full-Frame Image \textsf{eleanor} lightcurves from Sectors 1 through 54 \citep{2019PASP..131i4502F}. Details of the process are presented below. Thanks to this ongoing effort, we have already detected hundreds of thousands of TESS EB candidates \citep{2021tsc2.confE.163K}, hundreds of triple and quadruple systems
\citep[e.g.][]{2022MNRAS.510.1352B,2022MNRAS.510.1352B,2023MNRAS.521..558R,2022ApJS..259...66K}, the first and second eclipsing sextuple systems \citep{2021AJ....161..162P,2023MNRAS.520.3127Z}, and even a transiting circumbinary planet \citep{2021AJ....162..234K}.

Here we present the latest results from our work in the form of a new catalog of \Nquads uniformly-vetted quadruple star candidates from TESS. These are composed of two EBs in a 2+2 hierarchical configuration and are, in essence, a natural continuation of the \cite{2022ApJS..259...66K} catalog (hereafter K22). Similar to K22, all targets presented here have passed rigorous center-of-light motion tests indicating that both sets of eclipses are on-target and ruling out nearby resolved sources. The catalog provides the ephemerides, eclipse depths and durations, as well as additional information such as interesting features identified in the lightcurves or potential eclipse time variations. We note that this catalog only contains quadruple candidates where the two component EBs are unresolved and considered to originate from the same target. We deliberately exclude resolved, potentially co-moving pairs of binary stars on very wide orbits as these are unlikely to exhibit dynamical interaction on reasonable timescales for detection and are thus outside our scope of interests.

This paper is organized as follows. In Section \ref{sec:detection} we outline the detection of the quadruple candidates, followed by a description of their analysis in Section \ref{sec:analysis}. Section \ref{sec:catalog} presents the details of the catalog and we summarize our results in Section \ref{sec:summary}.

\section{Detection Methods}
\label{sec:detection}

To detect the quadruple candidates, we followed the procedures detailed in K22 which is outlined below for completeness. Briefly, members of the Visual Survey Group (VSG) \citep{2022PASP..134g4401K} performed visual inspection of the lightcurves of TESS EB candidates for the presence of additional features of interest such as extra eclipses/transits, unusual eclipse/transit shapes, coherent stellar variability, etc. The VSG is a dedicated team of expert citizen scientists specializing in surveying a large number of targets and has already contributed to major discoveries of transiting planets and multiple stars from NASA's Kepler and TESS missions. Altogether, VSG members have inspected millions of lightcurves to date and have co-authored dozens of peer-reviewed publications \citep{2022PASP..134g4401K}. 

To examine the lightcurves, the volunteers use personal computers, the LcTools software system \citep{2019arXiv191008034S,2021arXiv210310285S}, and other custom scanning and viewing software designed for interactive manipulation and analysis of photometric data. An example output from LcTools is shown in Fig. \ref{fig:LcTools}. Whether the lightcurves are viewed one at a time or by scanning batches of multiple files, years of experience have taught the volunteers how to quickly and reliably spot interesting features. The volunteers can typically inspect a lightcurve in less than ten seconds, in some cases much faster (K22). More details about the VSG detection methods and workflow can be found in \cite{2022PASP..134g4401K} and references therein. 

The EBs surveyed by the VSG citizen scientists were discovered in the Full-Frame Image TESS data by machine-learning methods applied to \textsc{eleanor-lite} lightcurves \cite{2022RNAAS...6..111P}\footnote{Sectors 1 to 6 available on MAST as High Level Science Products at the time of writing.}. The lightcurves were extracted with a local implementation of the \textsc{eleanor} pipeline \citep{2019PASP..131i4502F}, and the machine-learning procedure described in \cite{2021AJ....161..162P}. Briefly, we trained a neural network to identify light curves containing eclipses. This classification focused purely on the detection of eclipse-like features, with no periodicity requirement.  After constructing all light curves brighter than 15th magnitude in a {\em TESS} sector (usually ranging from $\sim$1 million to $\sim$10 million depending on density), we performed inference on the light curves using the neural network. The neural network could then significantly reduce the number of light curves requiring manual review by identifying binaries, which are then a much more likely source of triples and quadruples than random stars. The most promising candidates, generally in the range of the low tens of thousands (again depending on the density of stars in a given {\em TESS} sector), were manually reviewed by the VSG.  

\begin{figure}[h]
    \centering
    \includegraphics[width=0.99\linewidth]{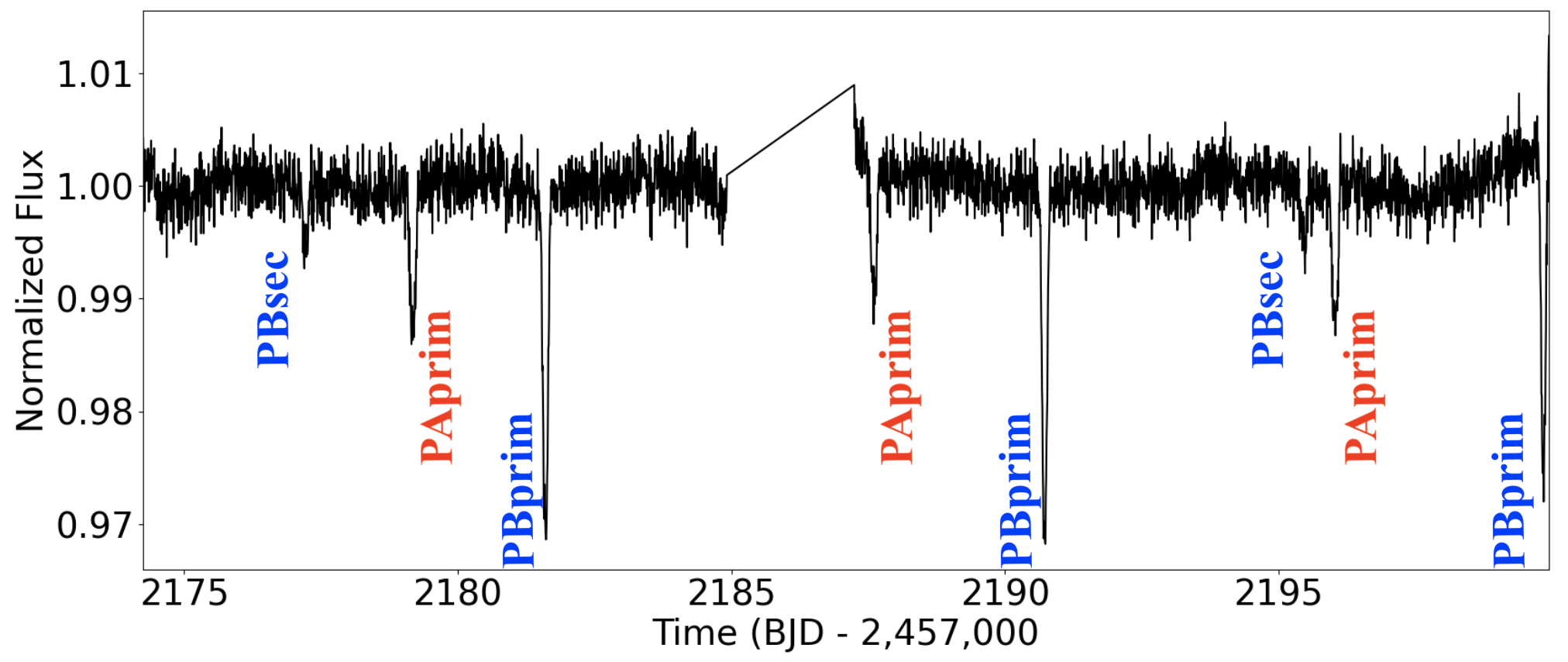}
    \caption{Full-Frame Image TESS \textsf{eleanor-lite} lightcurve from Sector 32 for quadruple candidate TIC 244279814 consisting of two eclipsing binary stars with PA = 8.42 days and PB = 9.12 days as labelled on the figure.}
    \label{fig:lc_example}
\end{figure}

For some targets, the VSG examined the publicly-available QLP \citep{2020RNAAS...4..204H} and short-cadence SPOC lightcurves \citep{2016SPIE.9913E..3EJ} as well. An example of a VSG-identified quadruple candidate is shown in Fig. \ref{fig:lc_example} for the case TESS Input Catalog (TIC) 244279814. The target exhibits two distinct sets of eclipses following two different periods, PA = 8.42 days and PB = 9.12 days, where PB produces both primary and secondary eclipses. Further analysis (described below) confirms that the target is the source of both EBs.  

\section{Analysis Methods}
\label{sec:analysis}

Contamination from nearby sources (be they resolved or unresolved, foreground or background) as well as non-astrophysical artefacts are ubiquitous in TESS data and every detected quadruple candidate needs to pass a series of vetting tests. Here we describe our methodology to address these issues, outline examples, and discuss known issues and caveats.  

\subsection{Pixel-by-pixel Vetting}

Once a quadruple candidate is identified by the VSG, it must pass rigorous vetting and validation tests before it is promoted as a genuine candidate (instead of a false positive, FP) and included in the catalog. Generally, FPs can be due to instrument and/or data processing systematics (e.g. momentum dumps, Solar System objects passing across the field of view), genuine astrophysical signals (e.g. resolved nearby EB and/or variable stars), or a combination of all three (see K22 for examples). The first step in ruling out such FPs is the pixel-by-pixel analysis described here. 

Due to the large pixel size of TESS (21 arcsec), most of the quadruple candidates found via the LcViewer application in LcTools are discarded as FPs that is simply due to a second nearby EB which is spatially resolved. We refer the reader to Figures 2 and 3 in \citep{2022PASP..134g4401K} for an example of such situation. As mentioned above, even if the nearby EB is co-moving with the target (according to Gaia DR3 measurements), such that the two potentially form a genuine wide quadruple system, the candidate is removed from consideration for this catalog. 

Overall, for a candidate to pass the pixel-by-pixel vetting, we utilize the interactive features of Lightkurve \cite{2018ascl.soft12013L}. As an example, TIC 144475902 was identified as a quadruple candidate in Sector 34 via LcViewer (Fig. \ref{fig:pixel_by_pixel_1}, upper panel). Using Lightkurve, we locate the pixel position of the target star from the Skyview image (Fig. \ref{fig:pixel_by_pixel_1}, lower panel; column 850.2, row 193.9) to select the proper aperture. Subsequently, we inspect the eight adjacent pixels individually, one by one, to confirm that the eclipses in the surrounding pixels scale equally in comparison to the eclipses in the chosen aperture (Fig. \ref{fig:pixel_by_pixel_2}). In such cases, a candidate will be promoted for photocenter vetting as described next. For TIC 144475902, the photocenter analysis confirmed that both sets of eclipses are on-target. 

\subsection{Photocenter Vetting}
\label{sec:photocenter_vetting}

Following the pixel-by-pixel vetting described above, we perform a detailed analysis of the center-of-light motion during the detected eclipses for both EBs of the quadruple candidate using the difference imaging method described in \cite{2019AJ....157..124K}. Briefly, for each eclipse of each binary, and for each sector, we first create the difference image by subtracting the corresponding average in-eclipse image (IE) from the average out-of-eclipse image (OOE). The latter is the average of the before-eclipse and after-eclipse images. The IE and OOE images have the same pixel size and are averaged over the same number of cadences, set by the measured eclipse duration. The measured center-of-light of the difference image represents the pixel position of the source of the detected eclipses. 

Next, to measure the center-of-light of each difference image, we fit it with a Point-Spread Function (PSF) and the TESS Pixel Response Function (PRF). This is done for completeness, as sometimes one of the functions may not fit the difference image well, and sometimes the other. While the goodness-of-fit is automatically reported by the fitting routine, in some cases the difference image can be dominated by systematic artifacts that prevent reliable fit -- even if the routine reports a successful fit. To account for this complication, we qualitatively evaluate the measured photocenter for each eclipse by visual inspection of the corresponding difference image and either accept the measurement or flag it as unreliable. 

Finally, we compare the measured average photocenter from all detected eclipses to the catalog pixel position of the target. If there are no statistical differences between the two, the candidate passes the photocenter test and the eclipses are considered to be on-target. Otherwise, the corresponding EB is marked as off-target and the quadruple candidate is flagged as a false positive\footnote{We note that one or both sets of the detected eclipse can be off-target, and can come from two unrelated sources, neither of which is related to the target star.}. 

As an example, for the case of TIC 144475902 described above there is a nearby field star (TIC 144475903, red arrow in Fig. \ref{fig:pixel_by_pixel_1}) that could potentially be a source of contamination. At a separation of $\approx$ 6.6(0.3) arcsec(pixels) and magnitude difference of ${\Delta T \approx 3.2}$ mag, TIC 144475903 is rather close to the target and almost bright enough to produce the PA eclipses (depth of $\approx0.03$, required ${\Delta T < 3.1}$ mag). As shown in Fig. \ref{fig:144475903}, there is a clear shift in the measured photocenters away from the position of TIC 144475903 and towards the position of TIC 144475902, ruling out the former as a potential source and confirming that they originate from the latter. 

For completeness, we note that the difference image should have a single, well-defined bright spot superimposed on an otherwise dark background. In practice, the difference image can be affected -- or even completely dominated -- by astrophysical (e.g. nearby variable stars) or systematic (e.g. background variability) ``noise'' which makes the corresponding photocenter measurements unreliable \citep[e.g.][]{2022PASP..134d4401K}. Fig. \ref{fig:bad_diff} shows examples of good-quality and of poorly-defined difference images. 

\begin{figure}[h]
    \centering
    \includegraphics[width=0.995\linewidth]{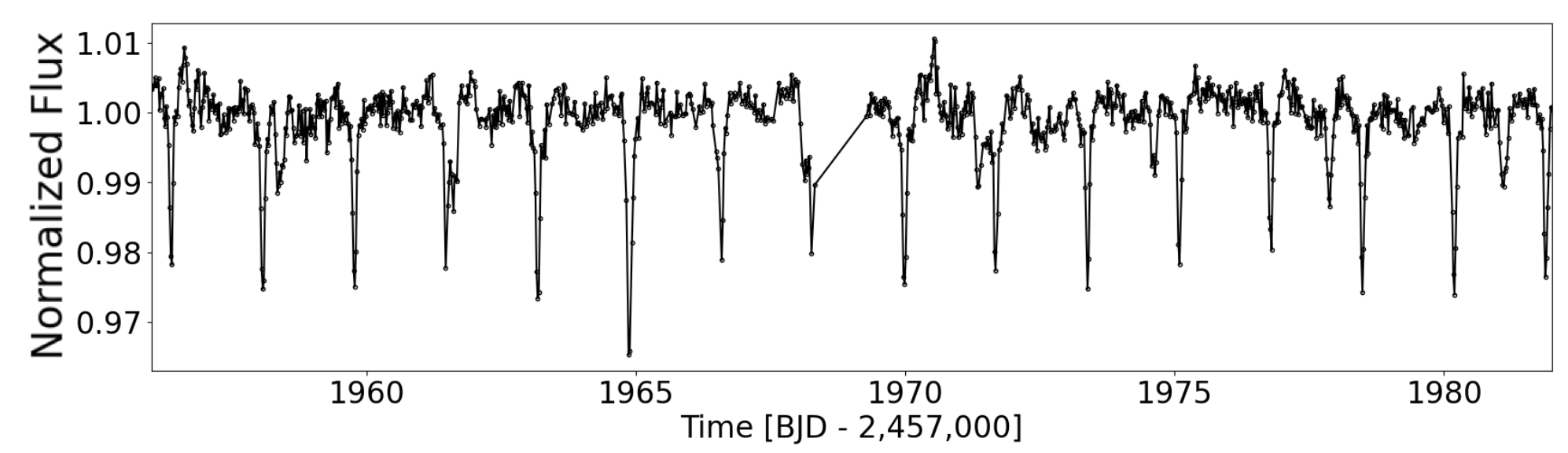}
    \includegraphics[width=0.995\linewidth]{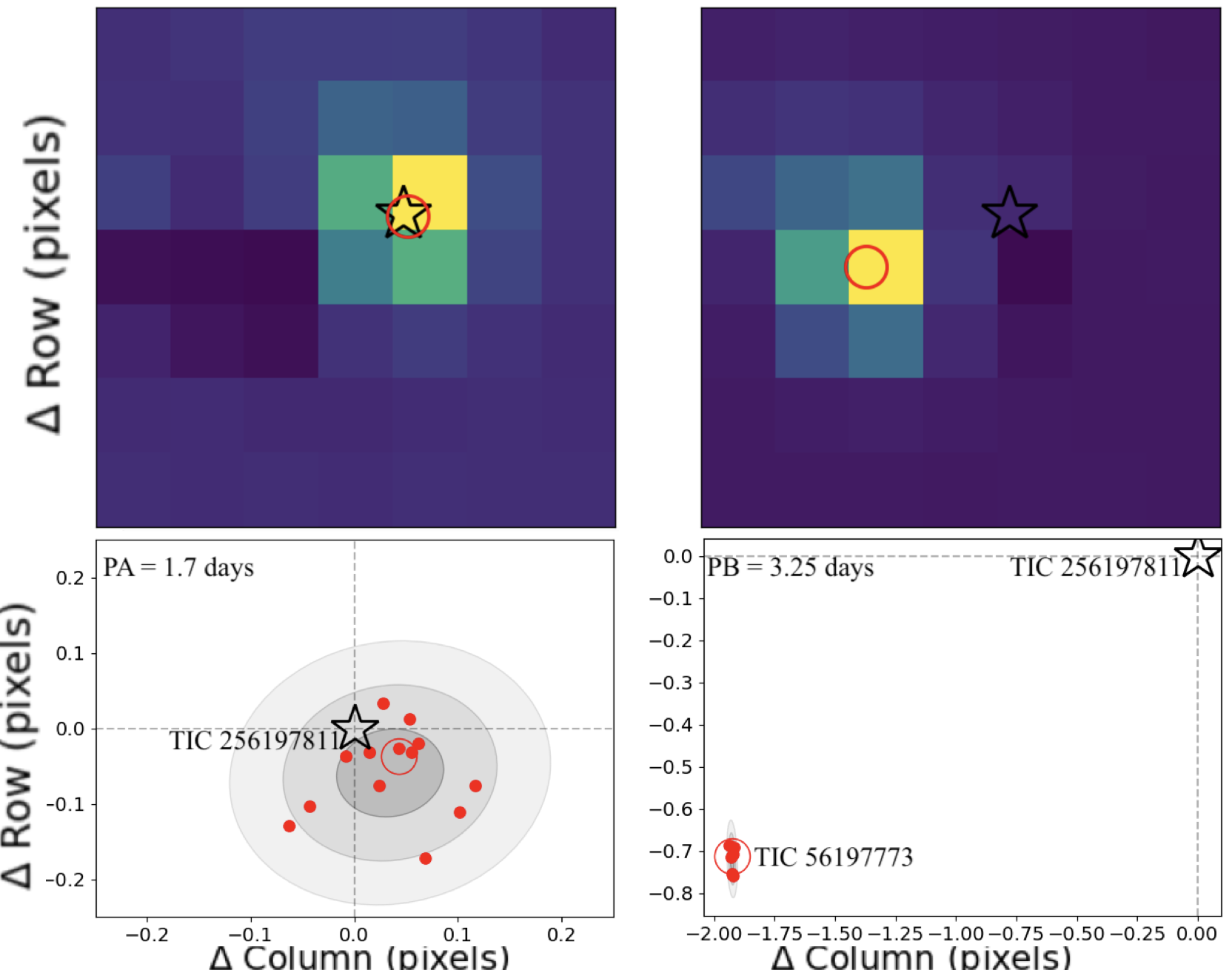}
    \caption{False positive quadruple candidate TIC 256197811. Upper panel: long-cadence \textsc{eleanor} lightcurve from Sector 24, showing two sets of eclipses with PA$\approx$1.7 days and PB$\approx$3.25 days, many of them overlapping. Middle row: The PA (middle) and PB (right) average difference images, along with the measured average photocenter (red circle) and the catalog position of the target (black star). Both represent a good-quality difference image and the respective photocenter measurements are reliable. Lower panels: Zoom-ins on the PA and PB difference images showing the individual measured photocenters for each eclipse (small filled circles), the average photocenter (large open circle), the corresponding 1-, 2- and $3-\sigma$ confidence intervals (grey contours), and the catalog position (black cross). We note that the two panels have dramatically different scales. The figure highlights a typical false positive scenario and illustrates the utility of the photocenter analysis.}
    \label{fig:FPCO}
\end{figure}

\begin{figure}[h]
    \centering
    \includegraphics[width=0.95\linewidth]{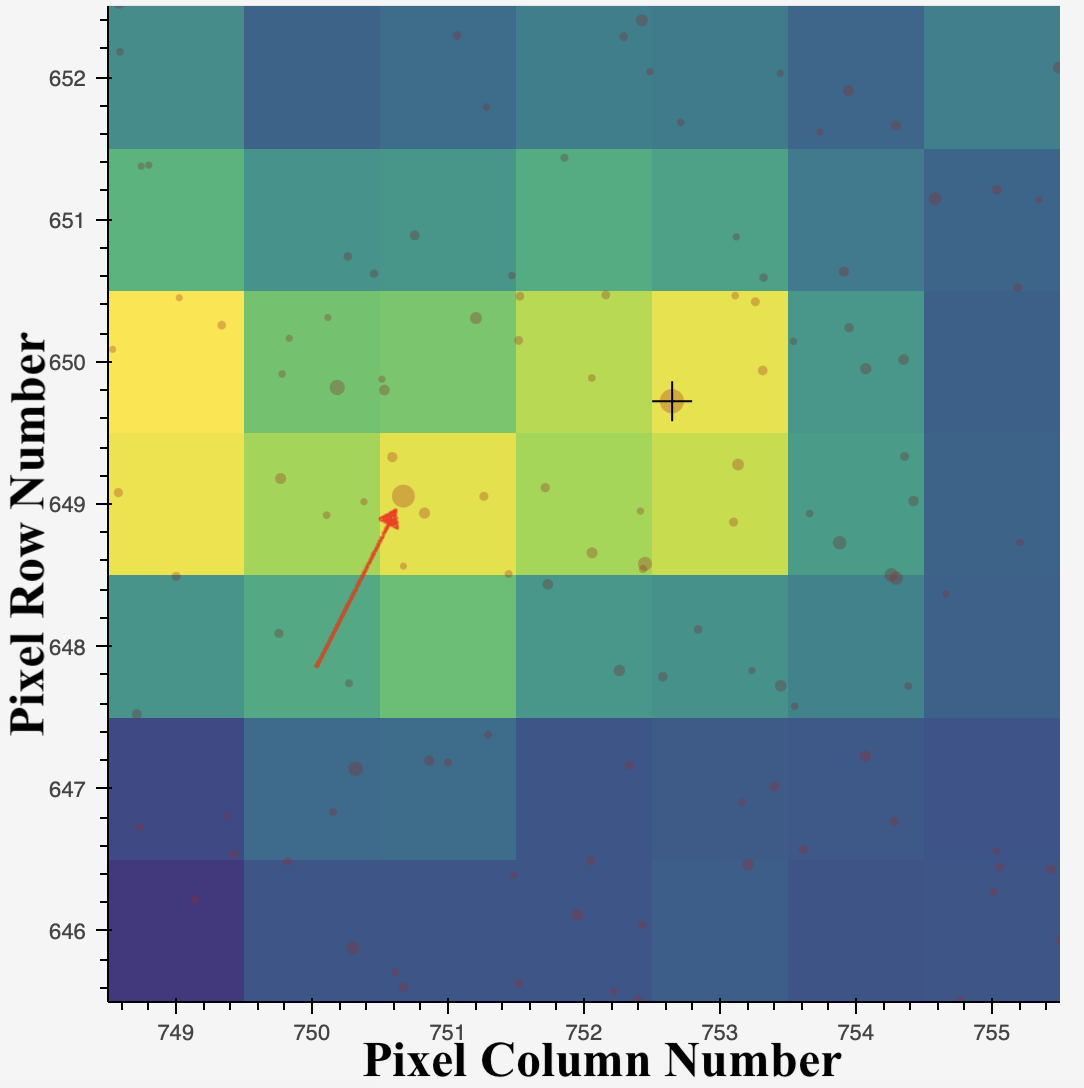}
    \caption{$7 \times 7$ TESS pixels Skyview image of the field around the target (black cross). The source of PB, TIC 56197773, is marked with a red arrow.}
    \label{fig:FPCO_SV}
\end{figure}

Fig. \ref{fig:FPCO} and \ref{fig:FPCO_SV} show an example of a quadruple false positive TIC 256197811 composed of two unrelated EBs, one on-target and the other off-target. The lightcurve shows two sets of eclipses with PA$\approx$1.7 days and PB$\approx$3.25 days, several of which partially or fully overlap. The measured photocenters show that the PA EB is on-target (lower left panel), while the source of the PB is an unrelated field star separated by about 2 pixels (TIC 56197773). We note that the 1-, 2- and $3-\sigma$ confidence intervals shown in the lower panels of Fig. \ref{fig:FPCO} as grey contours are calculated assuming that the individual measurements (small filled circles) follow a normal distribution. This assumption may or may not be valid, depending on the particular systematics. For more false positive examples, we refer the reader to K22. 

We note that we only measure the center-of-light for clean, unblended eclipses, as the difference imaging technique assumes that the OOE brightness of the target is not affected by additional events. Often, one or more eclipses can overlap, and are thus excluded from the photocenter analysis\footnote{Not to be confused with a syzygy in multiplanet systems, where the planets cross the line of sight with the same star. In contrast, overlapping eclipses in quadruple systems come from the two (unresolved) EB components.}. Sometimes, all eclipses are partially or fully blended -- i.e. there are effectively no clean eclipses for proper photocenter measurements -- and the difference imaging technique is not applicable. An example of this situation is shown in Fig. \ref{fig:VDLC} for the case of TIC 252301701 which exhibits two sets of eclipses following two nearly-identical periods, ${\rm P_1\approx1.487}$ days and ${\rm P_2\approx1.491}$ days (as measured by our disentangling method, described below). 

This is an unusual occurrence particular to eclipsing quadruple systems which, unlike transiting multiplanet systems, can have comparable periods and even ephemerides. For TIC 252301701, the ephemerides of the two EBs are such that the corresponding eclipses are slightly offset in Sector 9, but effectively fully-blended in S59. As a result, at first glance the S59 lightcurve looks like a single EB where the ``primary'' eclipse depth changes dramatically in a very short amount of time. After further scrutiny\footnote{And, naturally, a moment of disappointment.}, the slight asymmetry of the eclipses becomes apparent (thanks to the higher cadence compared to Sector 9) -- and even more so the ``emerging'' second set of eclipses near the end of the sector. In general, targets like TIC 252301701 are excluded from our catalog, even when there are no resolved nearby sources that can produce either set of eclipses (which is the case for TIC 252301701 as highlighted in the lower panel of Fig. \ref{fig:VDLC}). With that said, since the S59 lightcurve of TIC 252301701 effectively looks like a single EB it can, in fact, be used for difference image analysis. Indeed, the photocenter measurements confirms that the target is the source of the nearly fully-blended eclipses in S59, indicating that this is a genuine quadruple candidate. However, TIC 252301701 is not included in our catalog as it was independently discovered by \cite{2022MNRAS.517.2190R}.

\begin{figure}
    \centering
    \includegraphics[width=0.995\linewidth]{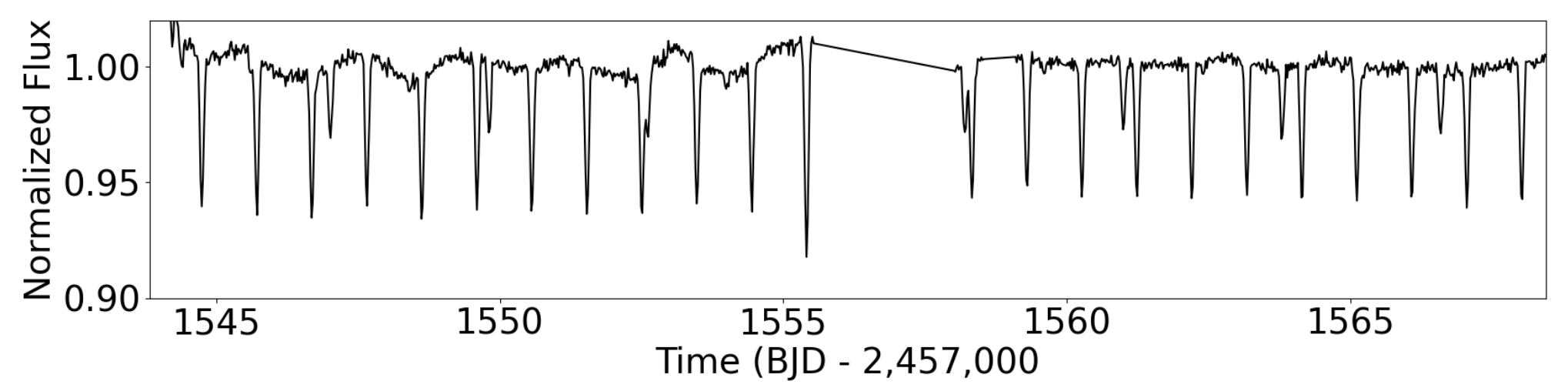}
    \includegraphics[width=0.95\linewidth]{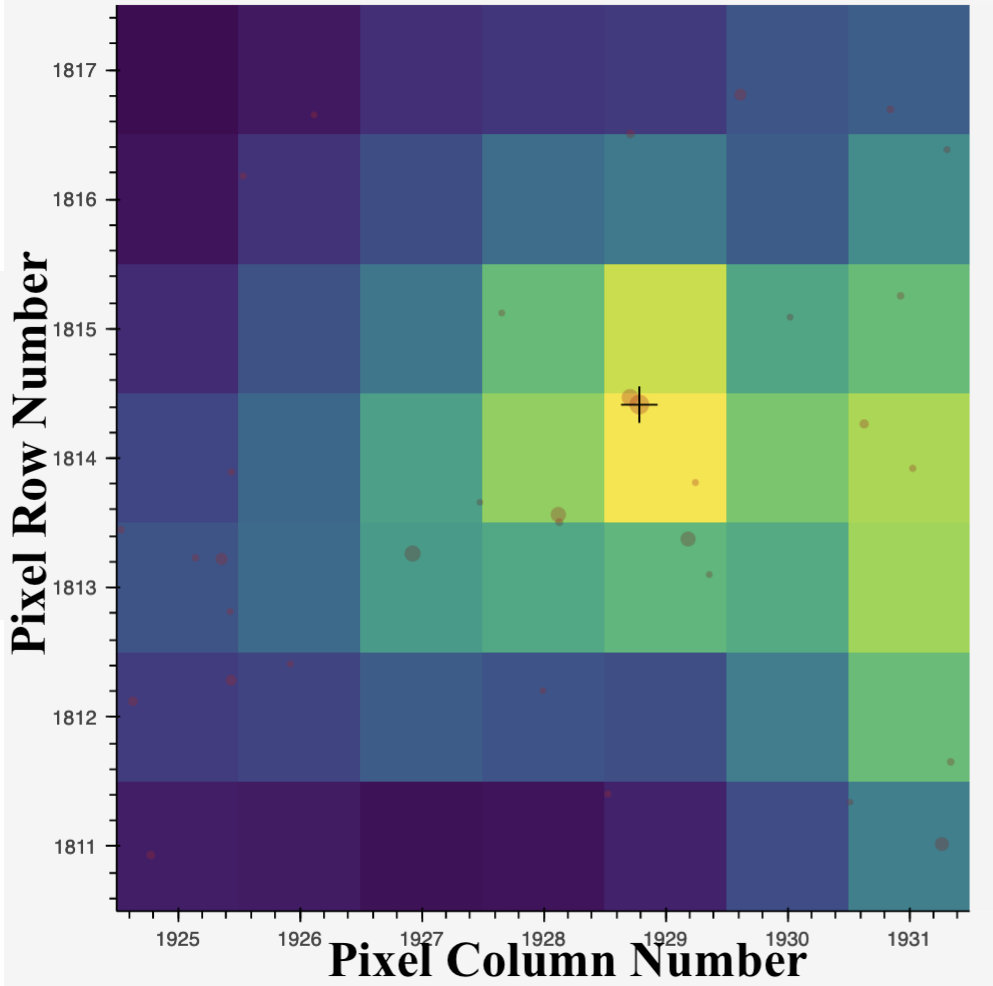}
    \caption{Quadruple candidate TIC 25712085. The upper panel shows the \textsc{eleanor} lightcurve from Sector 9, exhibiting two distinct sets of eclipses. The lower panel shows a $7 \times 7$ TESS pixels Skyview of the field centered on the target (Gaia 5439709915863478272), showing the nearby star TIC 873467932 (Gaia 5439709915863478144) which is bright enough to produce either set of eclipses (${\rm \Delta T = 1.1 mag}$) and too close to the target for reliable photocenter measurements (separation 2.2 arcsec, or 0.11 pixels). As a result, the candidate is marked as unclear and is excluded from our catalog.}.
    \label{fig:FSCP}
\end{figure}

Finally, there often are resolved field stars (either from Gaia or from other observations) that are bright enough to produce the detected eclipses as off-target false positives -- yet are too close to the target for reliable photocenter measurements. An example of this situation is shown in Fig. \ref{fig:FSCP} for the quadruple candidate TIC 25712085 where the nearby TIC 873467932 has a magnitude difference of ${\Delta T\approx1.1}$ mag and a separation of 2.2 arcsec (${\approx0.1}$ pixels). In cases like this, we are unable to verify the true source of the two EBs as we can neither confirm nor disprove it, mark the quadruple candidate as unclear, and remove it from consideration for this catalog.

\subsection{Ephemeris, Eclipse Depth and Duration Determination}

\begin{figure}[h]
    \centering
    \includegraphics[width=0.995\linewidth]{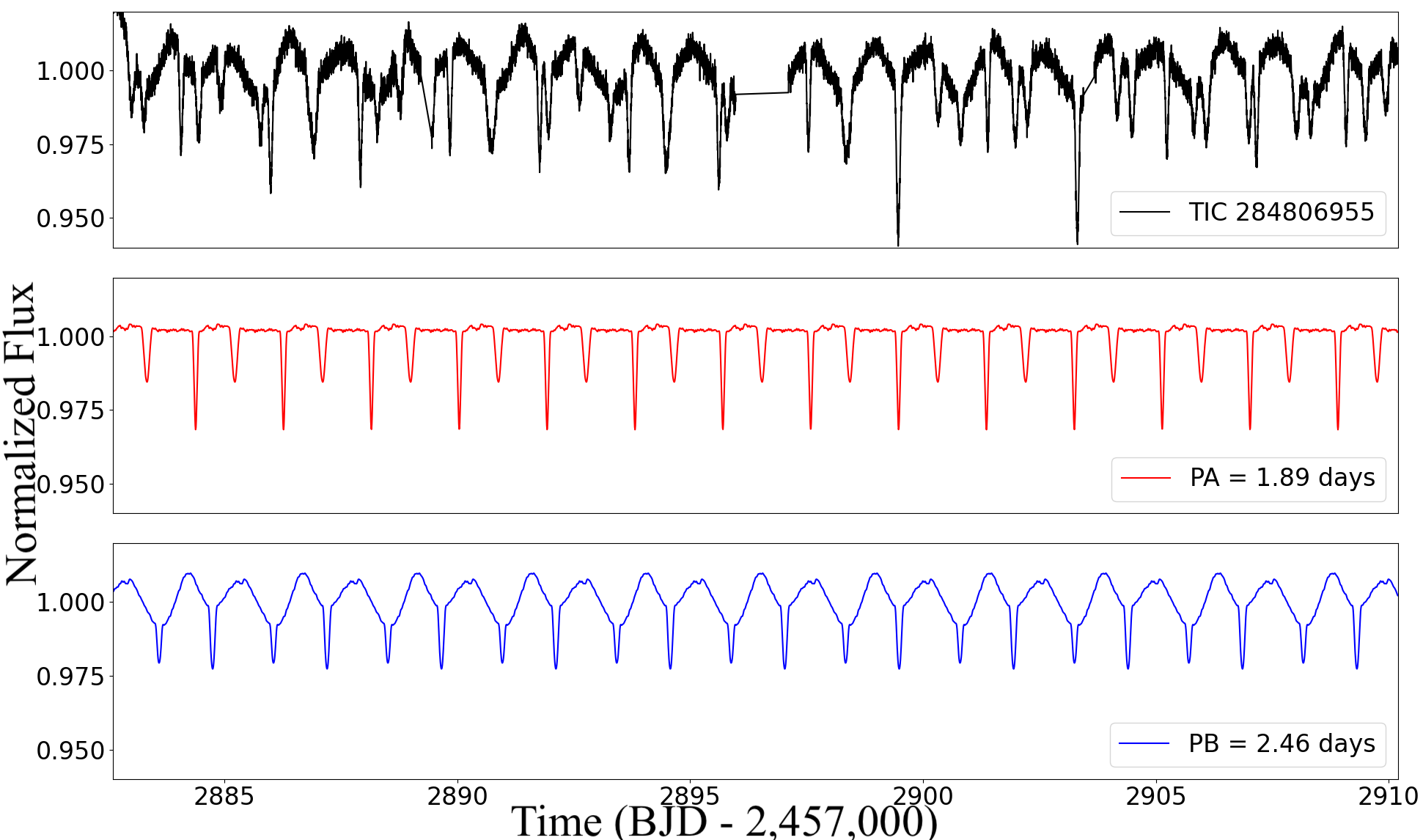}
    \caption{Fourier-disentangled lightcurve of quadruple candidate TIC 284806955, consisting of two EBs with PA = 1.89 days and PB = 2.46 days, as labelled on the figure. Upper panel: TESS lightcurve from Sector 58. Middle panel: Disentangled lightcurve for PA. Lower panel: same as middle panel but for PB.}
    \label{fig:disentangle}
\end{figure}

Many of the quadruple systems presented here have approximately commensurate periods and, consequently, produce multiple blended eclipses. To disentangle these systems and identify the properties of each component binary, we used a ``Fourier disentanglement'' approach similar to the method described in Section 3.1 of \citet{2021AJ....161..162P}. Briefly, the light curves for each quadruple system were separated by sector, and then simultaneously fit to 50 orbital harmonics, with a sine and cosine term for each, for both binaries' periods along with a constant offset. This resulted in a fit with 201 parameters for each individual sector, for each binary. The sector-by-sector approach was adopted to account for apparent differences in eclipse depths across sectors (caused by systematic effects) that would make simultaneous fitting difficult and most likely unreliable. An example of the method is shown in Fig. \ref{fig:disentangle} for the case of TIC 284806955 where the two EBs are heavily blended.

After disentangling, these light curves were plotted in quick-look plots that contained 
the light curves of both binaries and a comparison of the full quadruple light curve and the best-fit model, alongside with its residuals. Upon further inspection, there were several light curves that had issues when fitting due to one of two factors: outliers 
resulting from scattered light or data processing artifacts, and significant RMS scatter in the out-of-eclipse regions that made the eclipses hard to identify. We refit these light curves after examining them by eye and masking out points that were clearly affected by scattered light---especially during and after the orbit gap---and/or processing artifacts (which are usually present at the beginning and end of sectors). 

To measure the ephemerides, eclipse depths and durations of the quadruple candidates presented here, we used the methodology of K22. Specifically, we first applied the Box-Least Squares algorithm (BLS, \citep{2002A&A...391..369K}) to all available sectors stitched together and adopted the output as preliminary values. Where needed, we disentangled the lightcurve as described above, manually removed eclipses dominated by systematics as well as partial eclipses occurring near data gaps, and, for targets exhibiting strong out-of-eclipse modulations, detrended the data with a Savitsky-Golay filter. The latter is performed by first removing the eclipses from the lightcurve and then applying the filter, typically using 7th or 9th order and a window length of 121. Next, we fit each eclipse of each binary with four different functions: a Gaussian, a trapezoid, a generalized Gaussian (Eq. 1 in K22) and a generalized hyperbolic secant (Eq. 2 in K22). This approach allowed us to take into account the previously mentioned eclipse depth variations between different sectors and to refine the BLS ephemerides. The measured periods, eclipse times, depths and durations for the smallest chi-square fit among the four functions were adopted as the final values and reported in this catalog. To test for eclipse timing variations, we also fit a linear function to the measured eclipse times and evaluated the significance of any apparent deviations. 

We note that the workflow described above is an iterative process rather than a linear progression, with the various steps feeding into each other as needed. The process also evolved with time as new sectors and new versions of the \textsc{eleanor} pipeline became available during our analysis. Sometimes, new or updated data contradicted the initial interpretation and as a consequence some candidates were promoted, demoted, or even reclassified (e.g. from a triple to a quadruple or vice versa, as described below). 

\subsection{Comparison with Archival Data}

As noted in Section \ref{sec:photocenter_vetting} and highlighted in Fig. \ref{fig:FSCP}, quadruple candidates where potential contamination from Gaia-resolved nearby sources cannot be accounted for through photocenter analysis are excluded from our catalog\footnote{The only exception is TIC 387288959, as discussed below.}. For completeness, here we also utilize the available photometry from the ASAS-SN \citep{2017PASP..129j4502K}, ATLAS \citep{2018AJ....156..241H}, and Zwicky Transient Facility (ZTF, Decany et al. 2020) surveys to confirm that we see the same eclipses as in TESS data. Thanks to the better angular resolution and smaller pixel scale compared to TESS (8 arcsec for ASAS-SN, 1.86 arcsec for ATLAS, and 1 arcsec for ZTF), the lightcurves from these surveys are generally less affected by contamination from nearby sources -- which can be particularly problematic for TESS targets in crowded fields. Additionally, while these surveys do not have the comparably higher observational cadence and nearly-continuous coverage of TESS, they can provide a much longer observational baseline. Thus recovering a TESS-detected quadruple candidate in one (or more) of these surveys practically confirms that the corresponding eclipses are on-target.

We note that not finding the TESS eclipses in ASAS-SN, ATLAS, or ZTF does not necessarily rule out the candidate. The archival observations often contain a relatively small number of data points (e.g., in the typical range of 1000 to 3000 points), a signal-to-noise ratio that is relatively low, missed the eclipses, or the target was not observed at all. Additionally, ASAS-SN is saturated for targets brighter than 11th magnitude, and ZTF is saturated for targets brighter than 13th magnitude (priv. comm.). Thus recovering the eclipses in archival photometry is not always feasible.

As an example, the pubic ZTF repository provides lightcurves for 53 out of the 101 quadruple candidates presented here. Of these 53, we easily recovered both sets of eclipses for 30 targets (as noted in the catalog presented below). One set of eclipses is also easily recovered for 13 additional targets while the signal-to-noise and/or sampling of the ZTF data is insufficient for the robust recovery of the second set of EB eclipses. For three targets (TIC 779824, TIC 270360534, and TIC 370274336) one set of eclipses is recovered but the other set cannot be recovered because the corresponding EB period cannot be uniquely determined from TESS due to large data gaps (discussed below) or from the archival data. Finally, for 7 targets we couldn't recover either set of eclipses due to the same issues. Two of these 7 targets, TIC 219006972 (confirmed compact quadruple) and TIC 285853156, exhibit dramatic ETVs (amplitude of ${\rm \sim 0.1}$ days, see below) which make archival recovery highly challenging.

\begin{figure}
    \centering
    \includegraphics[width=0.48\linewidth]{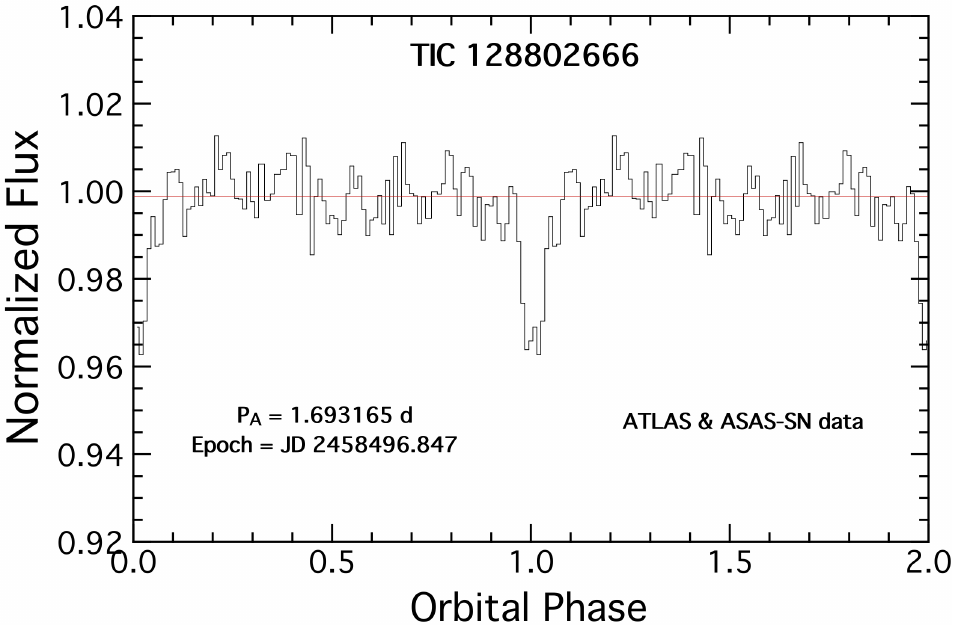}
    \includegraphics[width=0.48\linewidth]{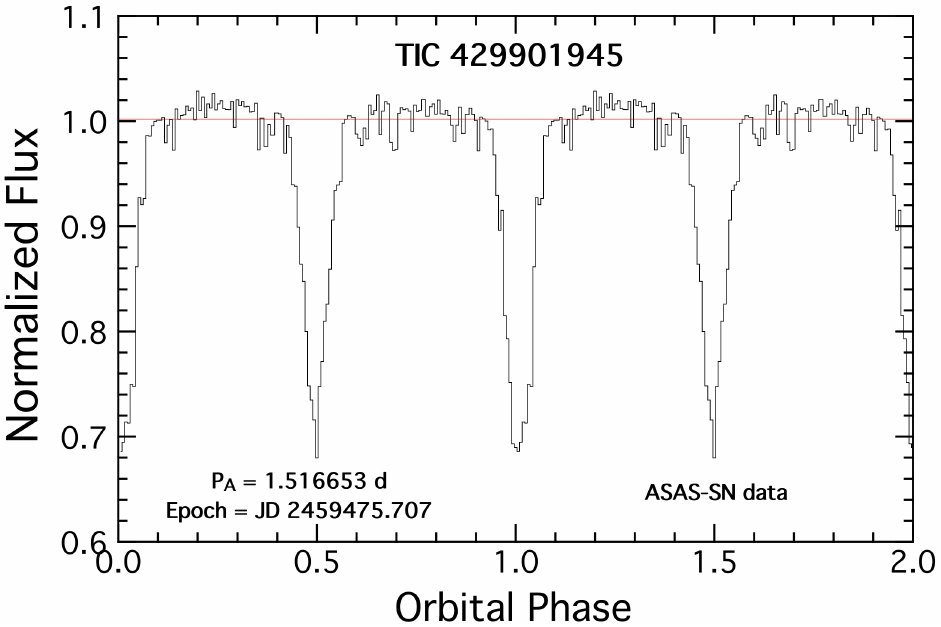}
    \includegraphics[width=0.48\linewidth]{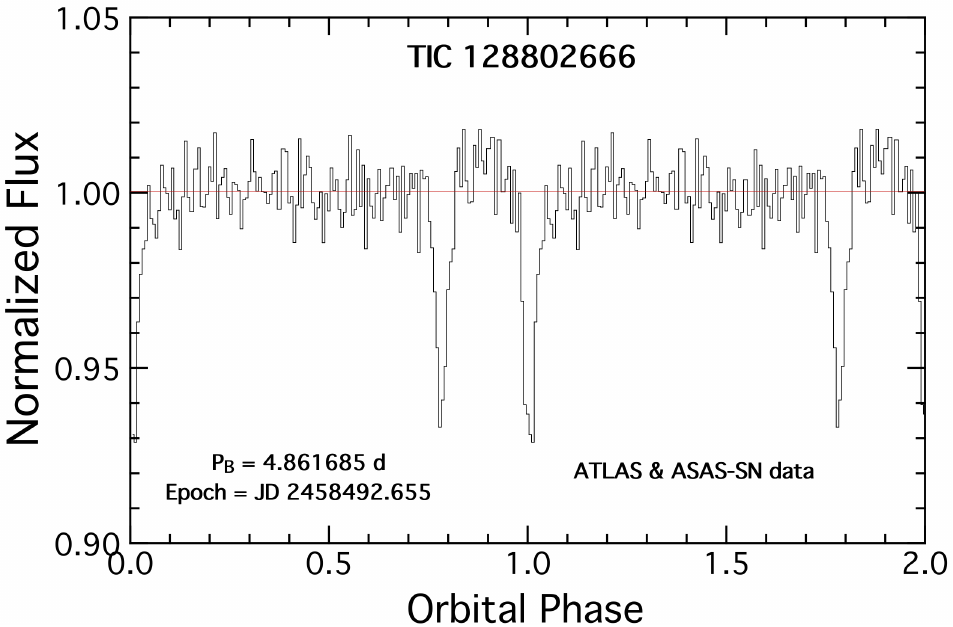}
    \includegraphics[width=0.48\linewidth]{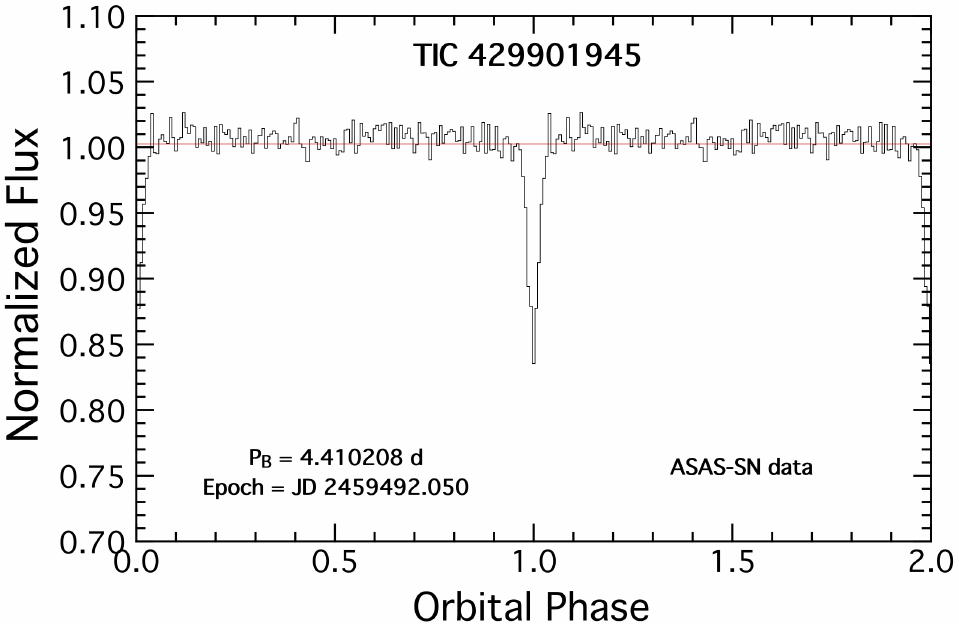}    
    \caption{Left panels: Phase-folded ASAS-SN and ATLAS lightcurves for binary A (upper panel) and binary B (lower panel) of TIC 128802666. Right panels: Same as left panels but for TIC 429901945. Both binaries are clearly detected in the data, confirming they originate from the respective target star.}
    \label{fig:ATLAS_archival}
\end{figure}

\begin{figure}
    \centering
    \includegraphics[width=0.48\linewidth]{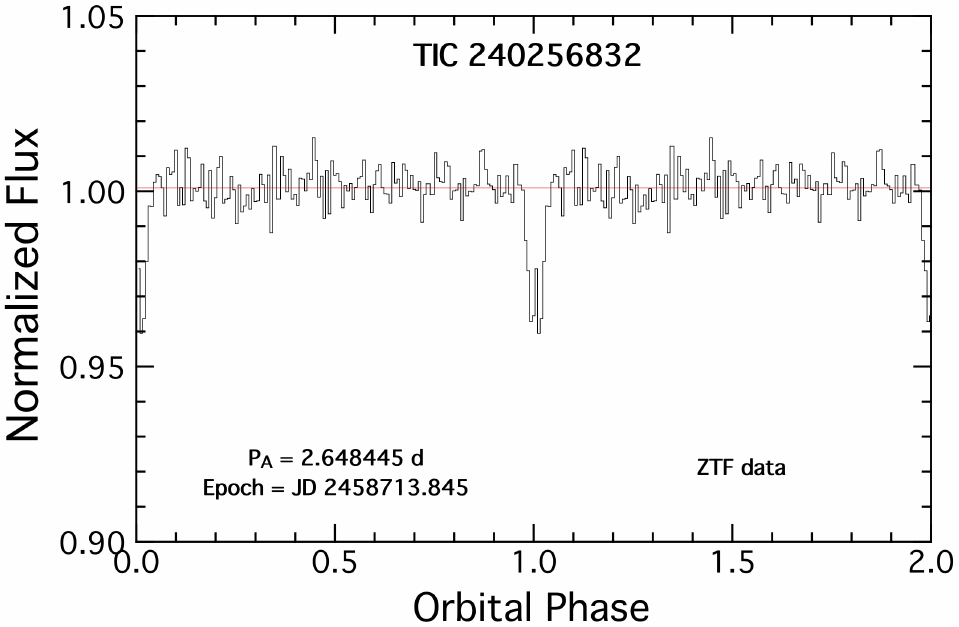}
    \includegraphics[width=0.48\linewidth]{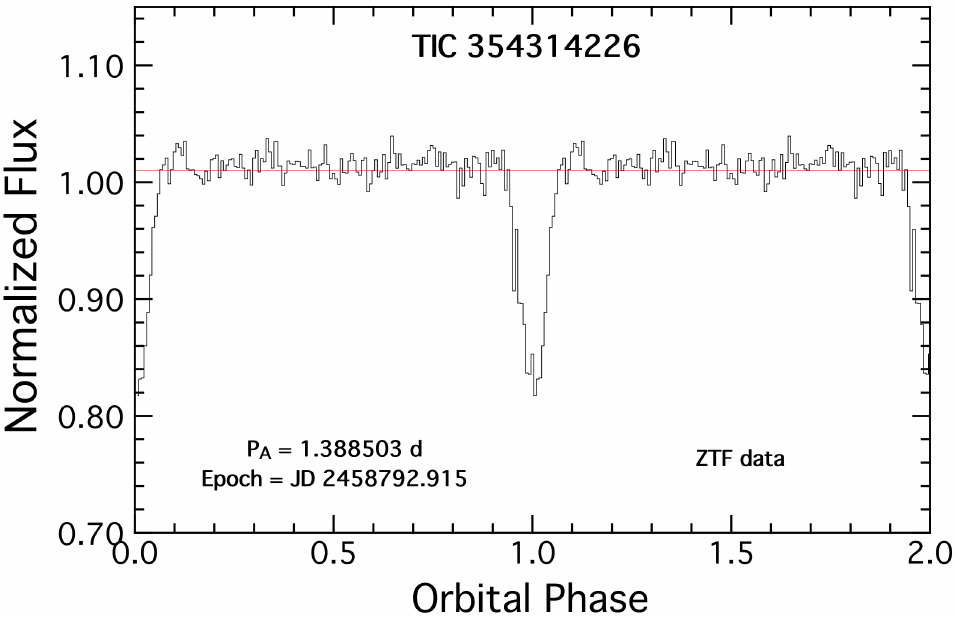}
    \includegraphics[width=0.48\linewidth]{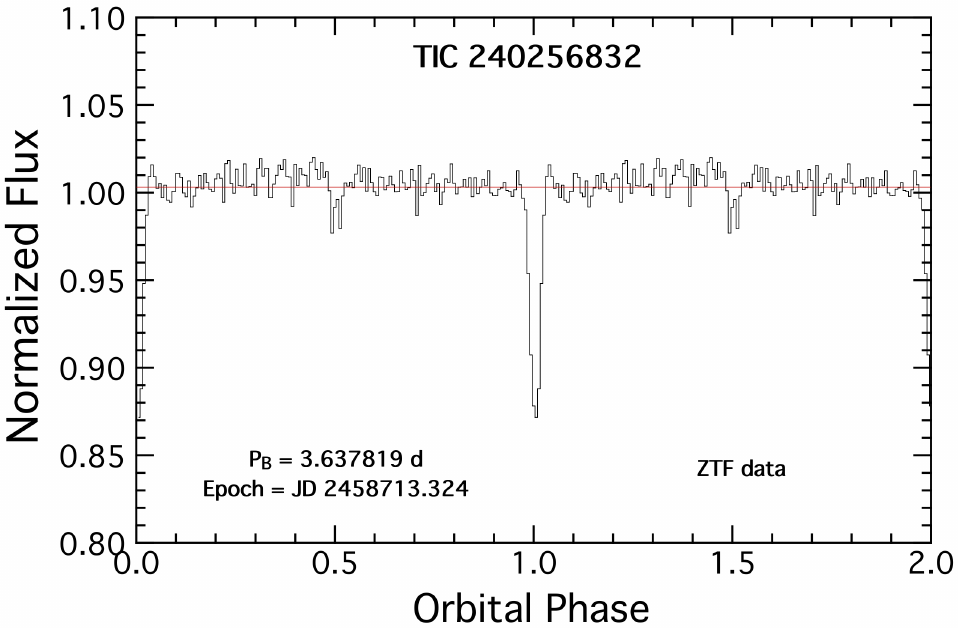}
    \includegraphics[width=0.48\linewidth]{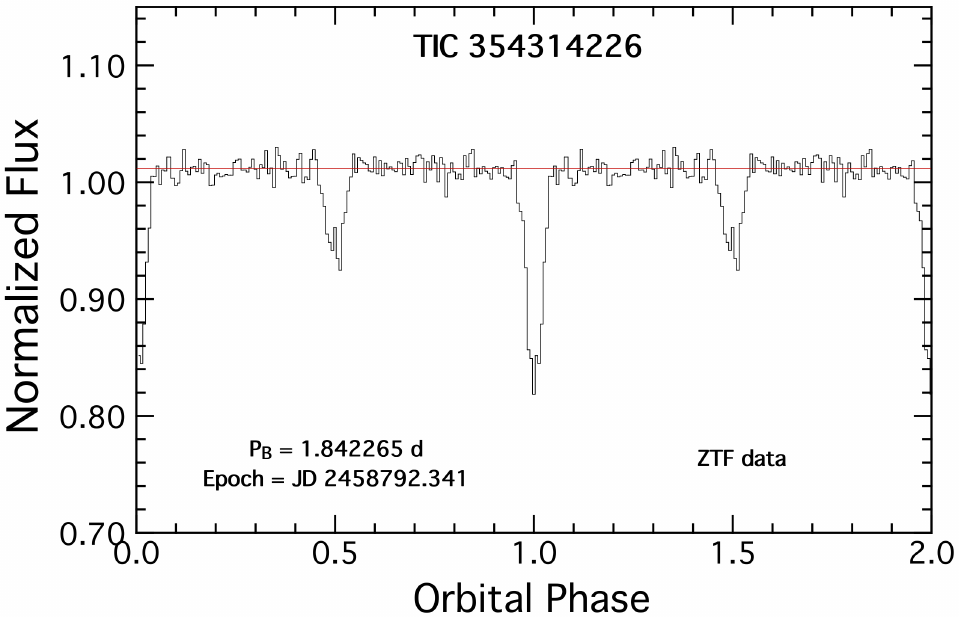}
    \caption{Phase-folded ZTF lightcurves for binary A (upper panels) and binary B (lower panels) of TIC 240256832 (left panels) and TIC 354314226 (right panels). Both binaries are clearly detected in the ZTF data for both targets.}
    \label{fig:ZTF_archival}
\end{figure}

To demonstrate the utility of cross-checking the photometry from TESS with that from surveys such ASAS-SN, ATLAS, and ZTF, we show a few illustrative cases in Fig. \ref{fig:ATLAS_archival} and \ref{fig:ZTF_archival}. The former shows the phase-folded ATLAS and ASAS-SN lightcurves for quadruple candidates TIC 128802666 and 120911334, while the latter shows the phase-folded data from ZTF for TIC 240256832 and TIC 354314226. In all cases, both EBs are clearly detected in the archival photometry, confirming that the respective target star is their source.

\section{The Catalog}
\label{sec:catalog}

The catalog of \Nquads eclipsing quadruple candidates presented here is provided to the community in a table format (see Table \ref{tbl:main_table}) including the TIC ID, the measured ephemerides, secondary phases, eclipse depths and durations for both sets of detected EBs of each quadruple candidate. The periods are labelled as PA/PB for the shorter/longer EB, respectively. For completeness, the table also lists the TESS magnitude, the Gaia DR3 identifier, composite effective temperature, and parallax, as well as additional comments relevant to the target. All quadruple candidates are labelled with the corresponding TESS/Goddard/VSG (TGV) identifier as well, following the nomenclature of K22 and starting from TGV-98. 

\begin{figure}[h]
    \centering
    \includegraphics[width=0.49\linewidth]{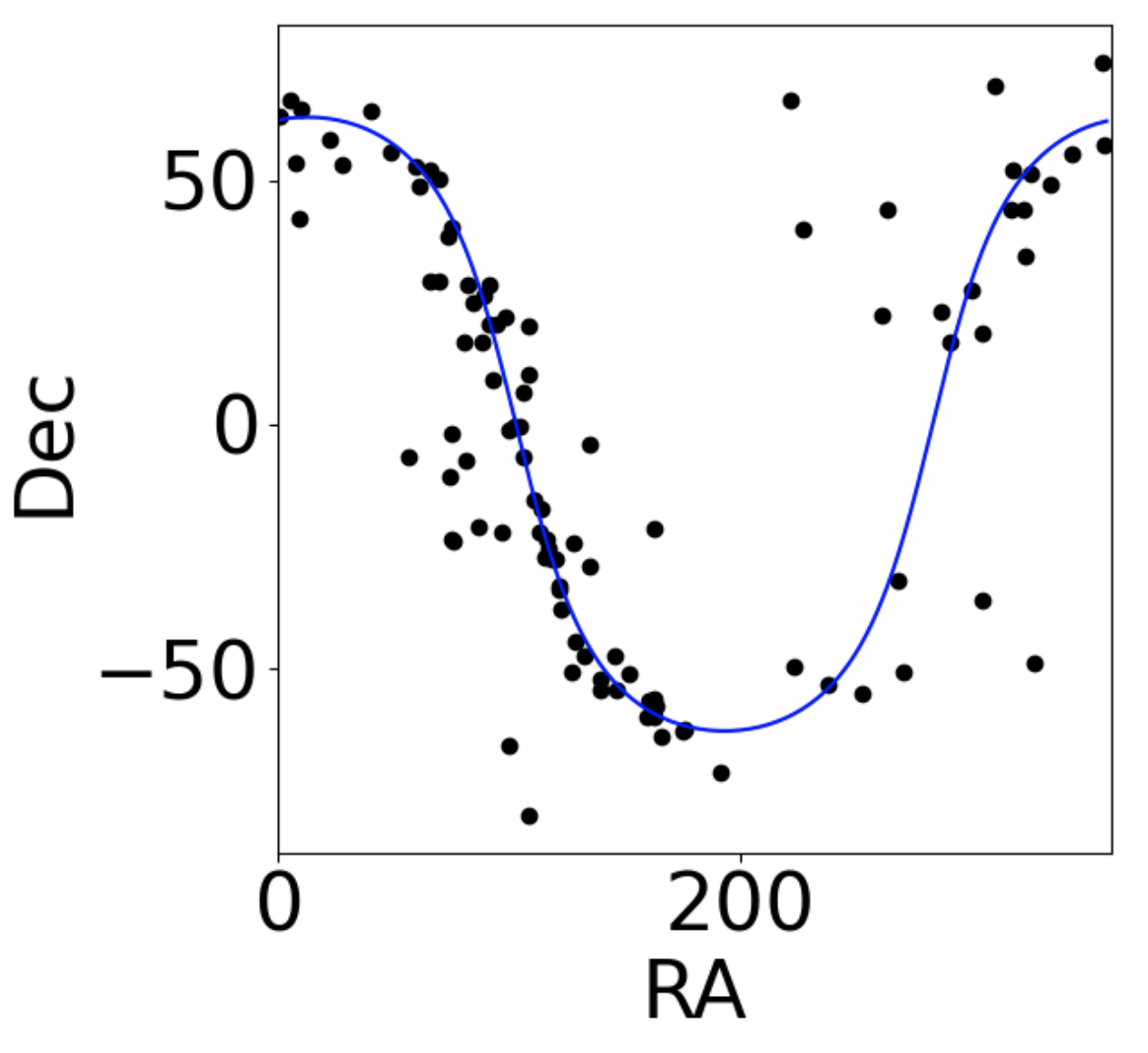}
    \includegraphics[width=0.49\linewidth]{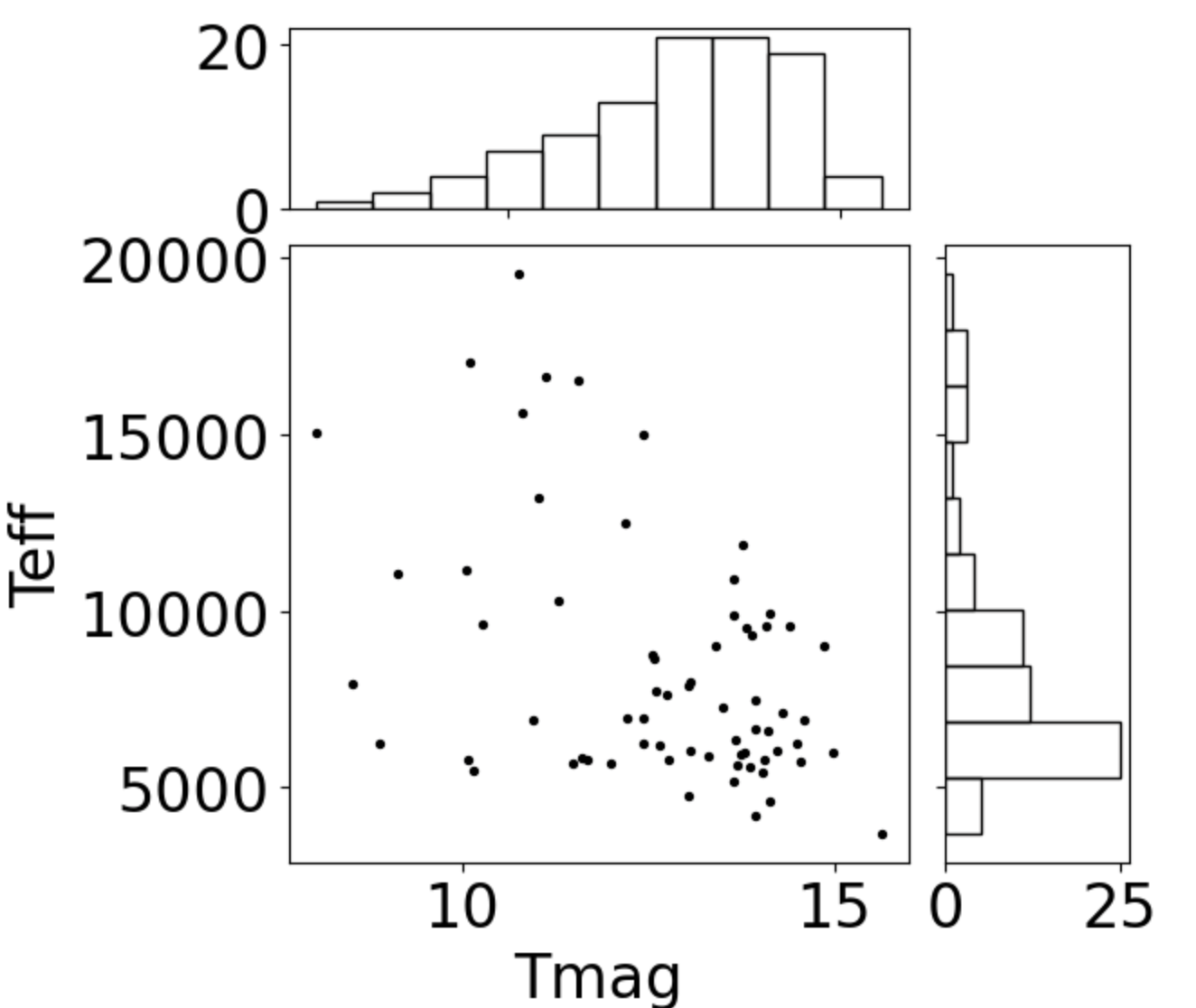}
    \includegraphics[width=0.99\linewidth]{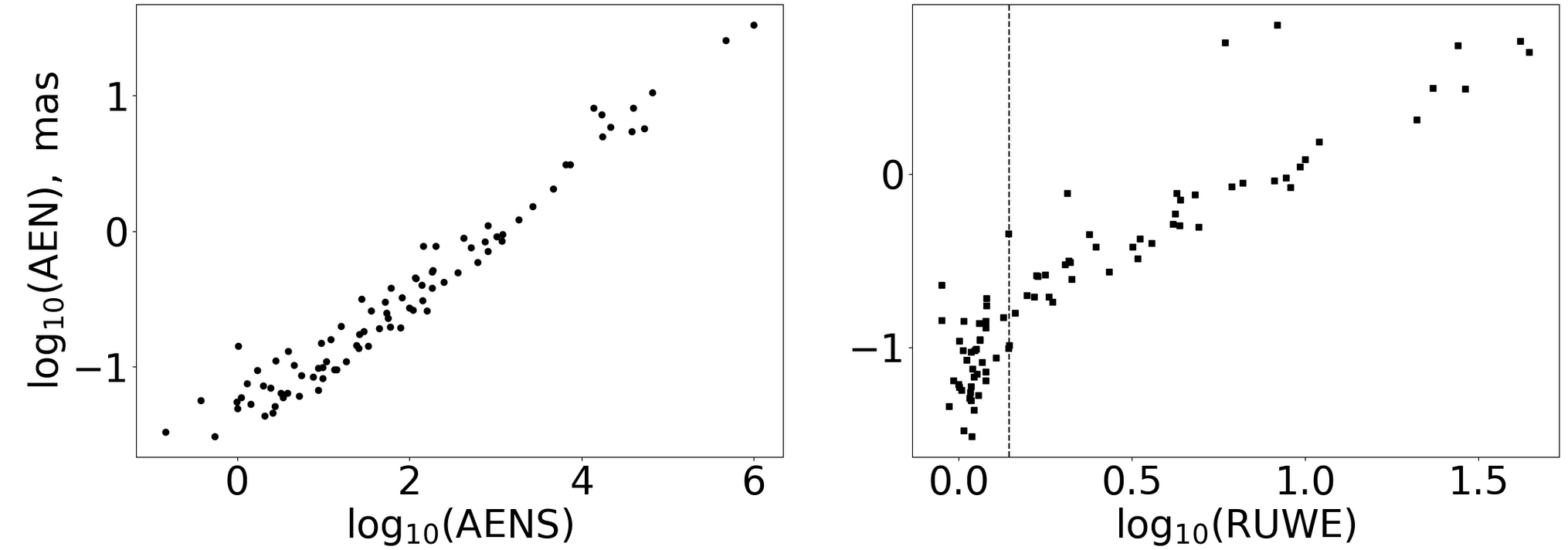}
    \caption{Upper panels: Sky position (left panel) and composite effective temperature (as provided by Gaia) as a function of the TESS magnitude (right panel) for the \Nquads quadruple candidates presented here. The blue line in the left panel represents the Galactic plane. Lower panels: The corresponding \texttt{astrometric\_excess\_noise} (AEN), \texttt{astrometric\_excess\_noise\_sig} (AENS), and renormalized unit weight error (RUWE) from Gaia DR3 (in log10 base). The dashed vertical line in the right panel represents RUWE = 1.4 which is typically considered as a potential indicator for unresolved companions.}
    \label{fig:basic_param}
\end{figure}

The sky position, TESS magnitude, and composite effective temperatures of the \Nquads quadruple systems, as provided by the TIC catalog and the Gaia archive, are shown in Fig. \ref{fig:basic_param}. The targets are spread across the sky, have a median magnitude of ${\rm T\approx12.84}$ mag, and a median (composite) effective temperature of ${\rm T_{eff}\approx6959}$ K. 

A comparison between the quadruple candidates presented here and archival databases can strengthen or contradict the interpretation based on the TESS data. Gaia DR3 data \citep{Gaia2021}, in particular, is a powerful resource to test for the presence of unresolved sources through the measured \verb|astrometric_excess_noise| (AEN), \verb|astrometric_excess_noise_sig| (AENS), and renormalized unit weight error (RUWE) as demonstrated by multiple studies \citep[e.g.][]{Belokurov2020,Penoyre2020,Stassun2021,Gandhi2022,2023MNRAS.523.2641R}. For example, a RUWE value greater than 1.4 can typically be an indicator for systems with unresolved companions \citep{Stassun2021}. 

The measured AEN, AENS, and RUWE for the \Nquads quadruple candidates are shown in Figure \ref{fig:basic_param}. 88 of the \Nquads systems have measured AEN $>$ 0, of which 16/9/3 have AEN $>$ 1/5/10 mas, respectively; the highest measured AEN -- about 33 mas -- is for TIC 470397849, a known EB. The measured AENS is greater than 3/5/100 for 72/67/37 targets, with a maximum of about 988,000 for TIC 470397849. The RUWE is greater than 1.4/5/10 for 43/16/7 targets, with a maximum value of 44.4 for TIC 274481742 (a known long-period variable). 

Altogether, these measurements suggest that many of the targets in our catalog may have unresolved companions and hint at the potential detection of orbital motion between the two EB components of the respective quadruple systems. For completeness, we have also noted targets with a non-zero Non-Single-Star (NSS) Gaia model. Five of the \Nquads systems have non-zero NSS: TIC 167800999, TIC 273919067, TIC 282005870, TIC 24700485, and TIC 356318101. All five have a RUWE between 1.6 and 11, as well as AENS between 78 and 2672, further strengthening the quadruple interpretation. TIC 273919067, TIC 24700485, and TIC 356318101 have acceleration solutions only, and therefore no outer period has yet been detected. TIC 167800999 and TIC 282005870 are further discussed below.

Interestingly, none of the \Nquads quadruple candidates presented here match with spectroscopic binaries listed in the \cite{2021AJ....162..184K} catalog based on APOGEE DR16 and DR17 data.  

\subsection{Period, Eclipse Depth and Duration Distributions}

The period ratios between the individual components of a hierarchical 2+2 quadruple system are important tracers of its formation and evolution. For example, Tremaine (2020) argued that resonant capture can produce systems with 3:2 and 2:1 period ratios between the two EBs, whereas period ratios of 1:1 are expected to be rare due to low capture efficiency \citep{2018MNRAS.475.5215B,2020MNRAS.493.5583T}. In a recent study of the distribution of period ratios in 72 2+2 quadruple candidates detected in OGLE data, \citep{2019A&A...630A.128Z} found a statistically significant excess of systems near the 1:1 and 3:2 period ratios. K22 presented a similar study, based on 72 systems discovered in TESS data, and found no evidence for enhancement at rational number period ratios. 

\begin{figure}[h]
    \centering
    \includegraphics[width=0.995\linewidth]{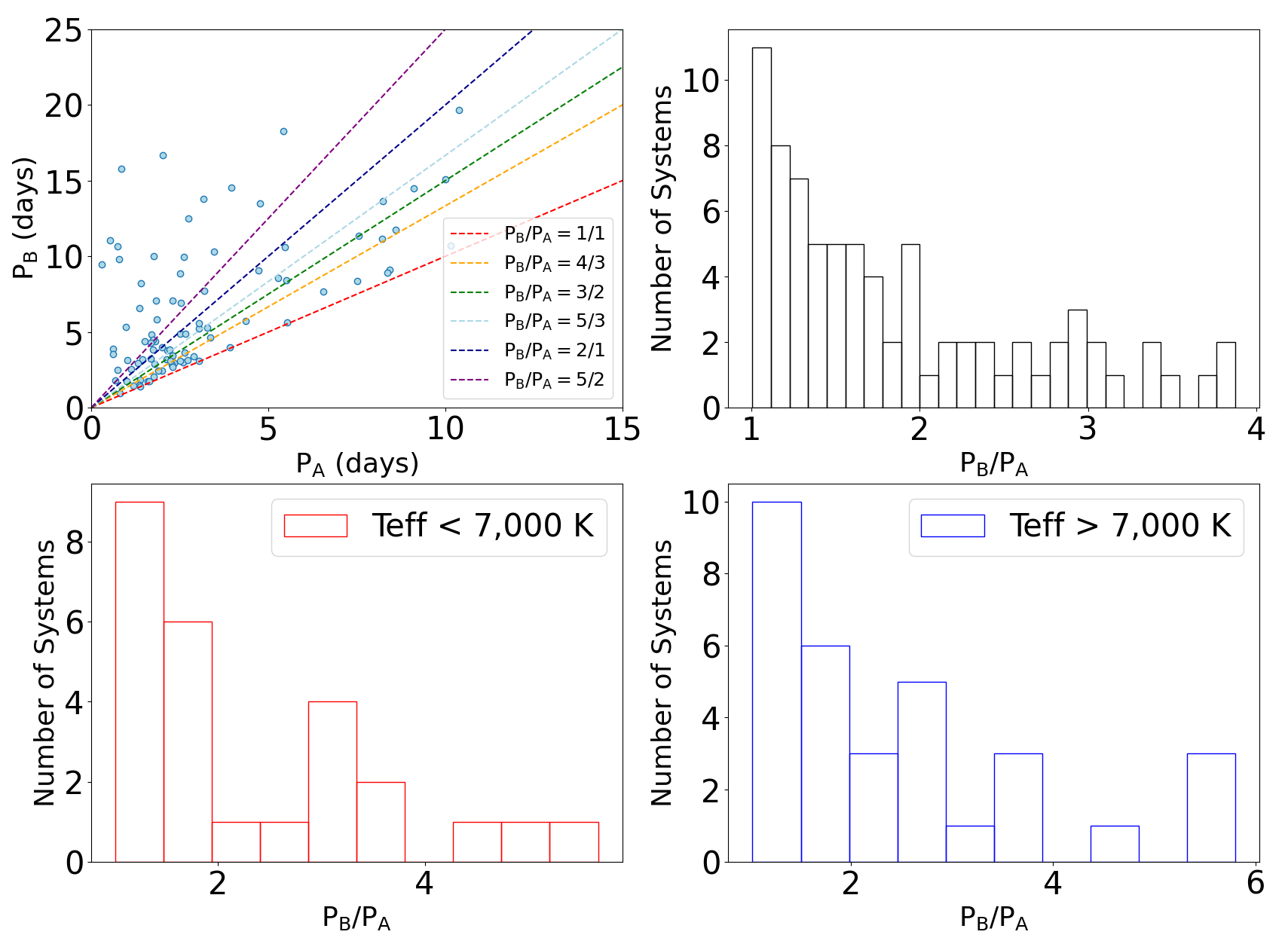}
    \includegraphics[width=0.995\linewidth]{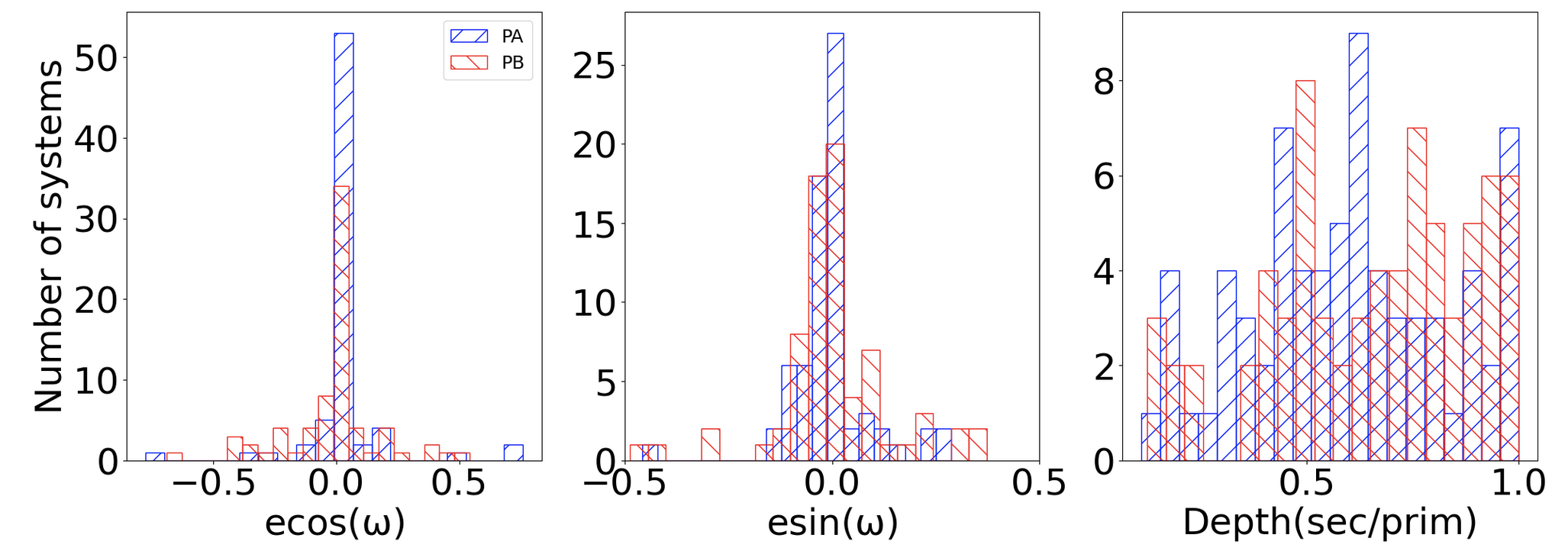}
    \caption{Upper left panel: Measured PA and PB periods of the component binaries of the \Nquads systems presented here (such that PA $<$ PB), along with several rational number ratios (upper left). Upper right panel: Corresponding period ratios PB/PA (smaller than 4, 81 out of the \Nquads systems). Middle panels: For comparison with Fig. 16 from Zasche et al. (2023). Same as upper right but for the subset of targets in our catalog with Gaia-provided ${\rm Teff < 7,000 K}$ (middle left) and ${\rm Teff > 7,000 K}$ (middle right). Lower panels: Calculated ${\rm e\cos(\omega)}$ (left) and ${\rm e\sin(\omega)}$ (middle), and measured eclipse depth ratios between secondary and primary eclipses for each binary of each quadruple system (right). Blue/red colors represent the values for the corresponding binary A/B, respectively.}
    \label{fig:periods_}
\end{figure}

The measured periods, in terms of PA and PB for each component binary, and the corresponding period ratios PB/PA for the \Nquads systems are shown in Fig. \ref{fig:periods_}. To facilitate a direct comparison with Zasche et al. (2019) and K22, the figure shows the systems with PB/PA $<$ 4 (81 out of \Nquads) and uses the same number of bins (26). While many of the systems seem to cluster near the 1:1 period ratio, none is at exact resonance. Only one system has a period ratio within less than 1\% of unity (TIC 358422952, PA = 3.0536 days, PB = 3.0757 days), and three other systems have period ratios within less than 4\% of 1:1 resonance (TIC 300987891, 459333241, and 459705607). Overall, we find no evidence for an excess of systems near rational number period ratios, in line with the results of K22.  

The calculated ${\rm e\cos(\omega)}$ and ${\rm e\sin(\omega)}$ for each binary of each quadruple candidate are shown in the lower panels of Fig. \ref{fig:periods_}. The values are close to zero for the vast majority of the systems, indicating the predominance of nearly-circular orbits. Fig. \ref{fig:periods_} also shows the measured depth ratios between the secondary and primary eclipses for each binary of each system. We find no compelling evidence for preferred values.  

\subsection{Discussion}

The catalog presented here is the product of a comprehensive search and analysis process yet it should not be treated as a complete set of eclipsing quadruple systems from TESS. Our machine-learning-powered search for EBs (which are the main focus of the subsequent visual inspection) was designed for efficient and reliable detection of detached systems rather than completeness. As such it likely missed a number of systems, in particular in the very short-period range of contact and overcontact binaries. 

In terms of overall numbers, at the time of writing members of the VSG have detected thousands of candidate stellar multiples from TESS, a large number of potential planet candidates, dipper stars, as well as other unusual astrophysical objects and events. Of these, to date we have identified 3038 targets as worthy of further investigation and analysis. The rest were discarded quickly after the initial detection due to obvious issues such as identical eclipse pattern as other targets, foreground contamination from Solar System objects, known systematics (e.g. momentum dumps), etc. Of the 3038 targets, 1758 have already been fully or partially vetted, including hundreds of clear and potential false positives. 

While some of the quadruple candidates included in this catalog have already been identified as eclipsing binaries, prior to TESS the majority were only known as single stars. During our analysis, we also independently detected a number of quadruple candidates listed in previous catalogs \citep[e.g.][]{2022MNRAS.517.2190R,2019A&A...630A.128Z,2022BAAS...54e.414E}, including several targets that we identified as false positives or potential false positives. Specifically, (i) TIC 280381105, 283690818, 362809861, 43250275, 50477999, and 63291675 show a clear photocenter offset during eclipses; (ii) TIC 465246981 has a resolved nearby star bright enough to produce one or both sets of eclipse, but is too close to the target for reliable photocenter measurements; and (iii) TIC 258837989 and TIC 19757900 show a single set of eclipses in the \textsc{eleanor} lightcurve.

\subsubsection{Individual Systems}
\label{sec:individual_systems}

Below we list several systems exhibiting potentially interesting features in addition to the two sets of detected eclipses. 

\begin{figure}[h]
    \centering
    \includegraphics[width=0.995\linewidth]{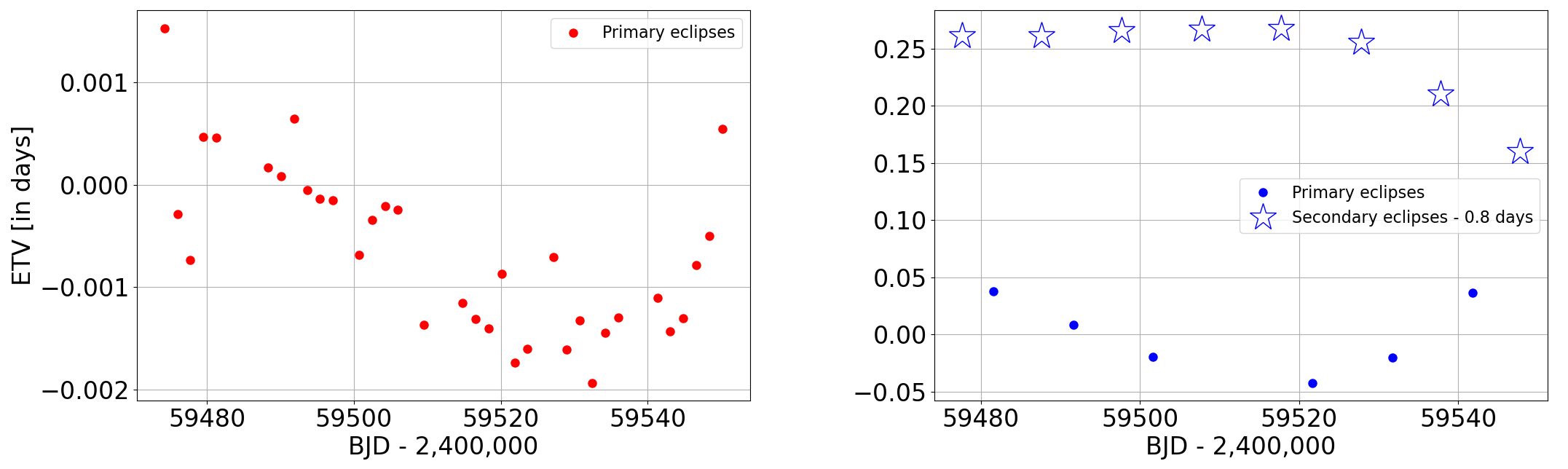}
    \caption{Measured eclipse timing variations for quadruple candidate TIC 285853156 (TGV-134) for the PA (left) and PB (right) components, in terms of Observed minus Calculated time as a function of the Observed time. The primary and secondary ETVs in the right panel are vertically offset by 0.8 days.}
    \label{fig:285853156}
\end{figure}

\begin{figure}
    \centering
    \includegraphics[width=0.995\linewidth]{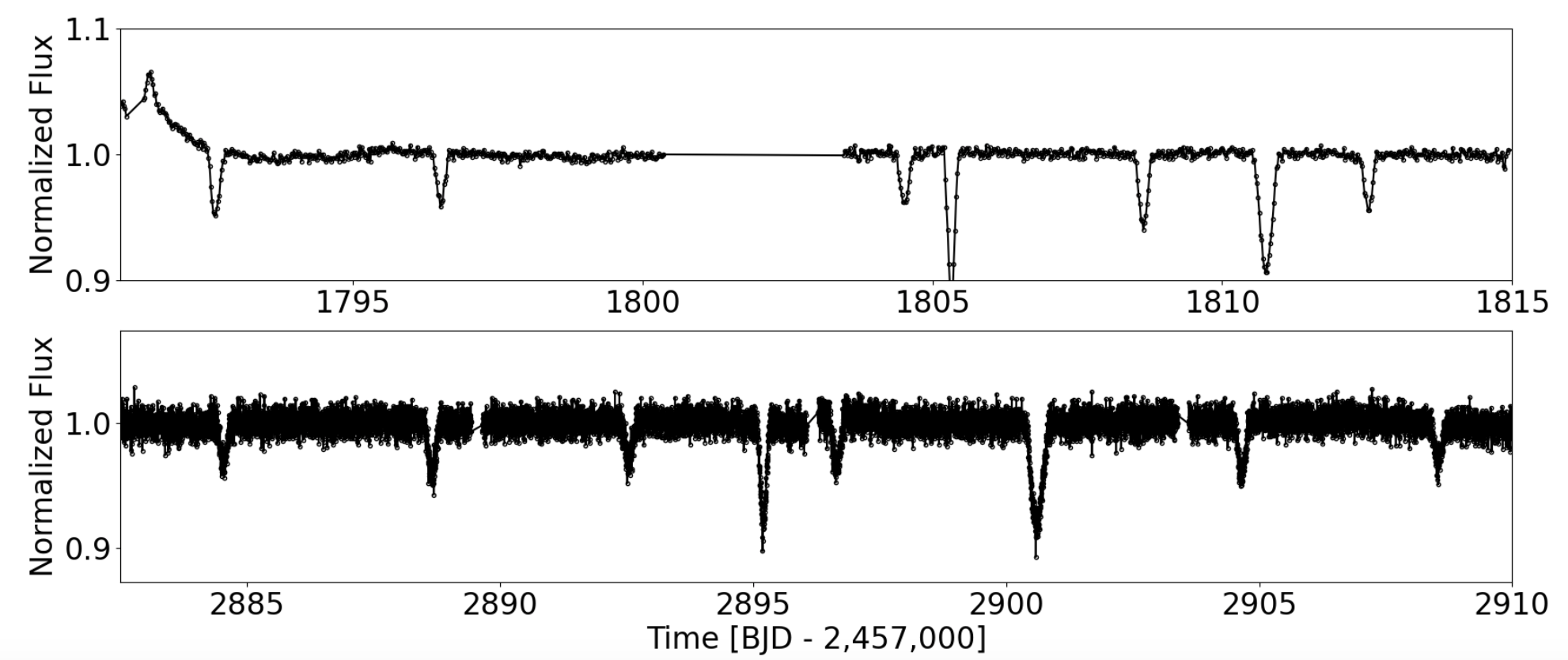}
    \caption{TESS lightcurve for quadruple candidate TIC 370274336. Upper panel shows data from Sector 18, lower panel shows data from Sector 58. One binary has an orbital period of PA = 8 days. The period for the other binary, PB is unclear as there are only two primary and two secondary eclipses separated by a large gap. With that said, PB must be longer than 20 days, ${\rm \approx1089.86/N}$ days where N is an integer, and the orbit quite eccentric.}
    \label{fig:370274336}
\end{figure}

\begin{figure}
    \centering
    \includegraphics[width=0.995\linewidth]{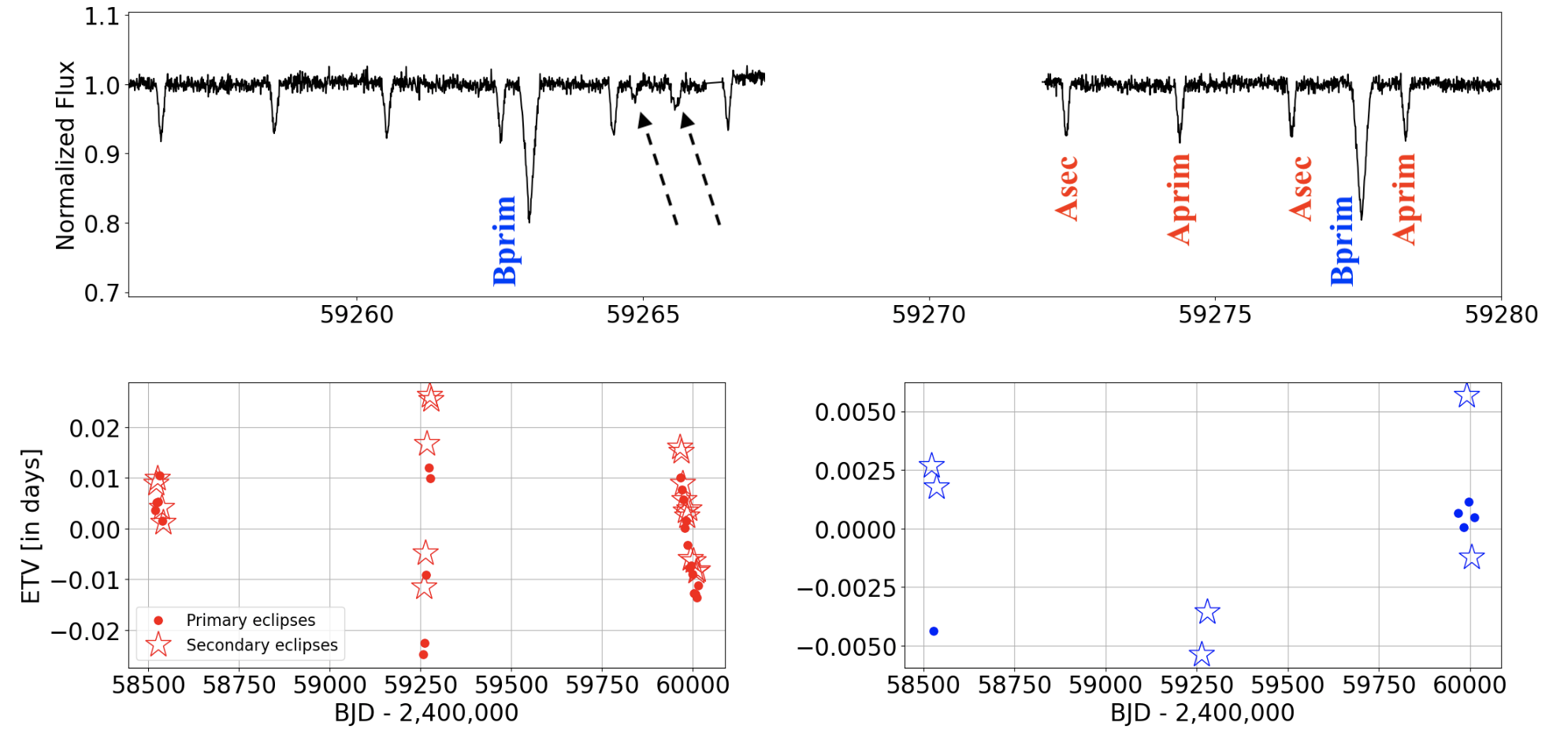}
    \caption{Upper panel: FITSH lightcurve of TIC 37376063 for Sector 35, showing the two clear EBs as well as the additional pair of events (dashed arrows) consistent with tertiary eclipses on binary A, suggesting a (2+1)+2 quintuple system. Lower panels: Same as Fig. \ref{fig:285853156} but for TIC 37376063. The PA binary (red) shows clear variations and there are potential non-linear deviations for PB (blue) as well.}
    \label{fig:37376063}
\end{figure}

\begin{figure}[h]
    \centering
    \includegraphics[width=0.995\linewidth]{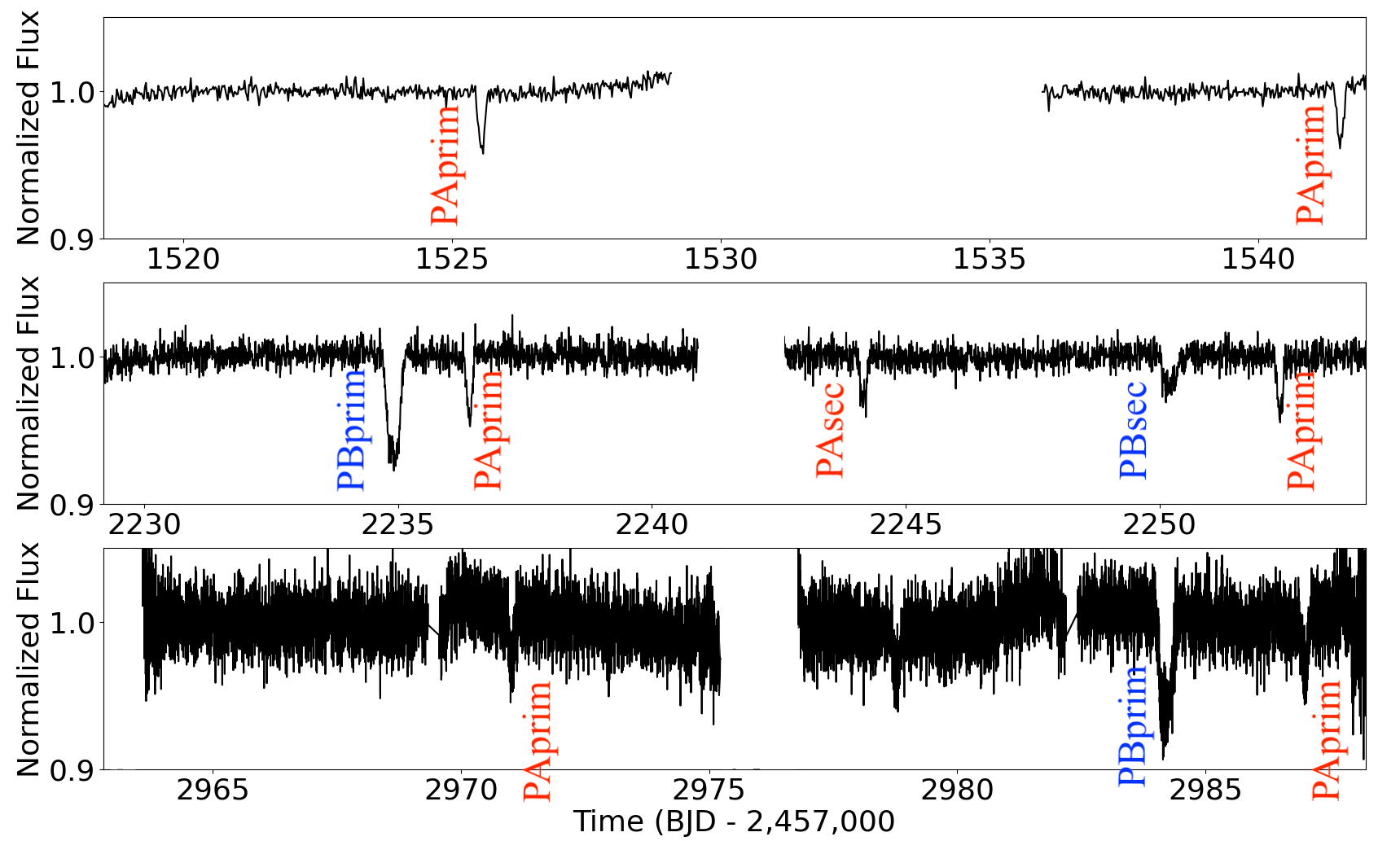}
    \caption{TESS lightcurve of TIC 779824 for Sectors 8, 34 and 61, showing one EB with a period of PA = 15.97 days (highlighted in red) and three additional eclipses, labeled in blue as PB, two primary and one secondary. Given the large gap between Sectors 34 and 61, PB must be ${\rm \approx749.29/N}$ days where N is an integer. }
    \label{fig:779824}
\end{figure}

\begin{description}

\item[$\bullet$ TIC 219006972 (TGV-120)]: A confirmed 2+2 quadruple system composed of two EBs with orbital periods of PA = 8.28 days and PB = 13.66 days, respectively. It was observed by TESS in Sectors 14, 15, 16, 21, 22, 23, 41, 48, and 49, and both EBs exhibit dramatic ETVs on a timescale of 168 days. The system is nearly co-planar with an outer eccentricity of about 0.25. At the time of writing, TIC 219006972 is the second shortest outer period confirmed quadruple system \citep{2023MNRAS.522...90K} after BU Canis Minoris (Pribulla et al. submitted). 

\item[$\bullet$ TIC 274862252 (TGV-98)]: The two component EBs have orbital periods of PA = 2.34 days and PB = 2.95 days, respectively, and many of the eclipses are blended. The target was observed in Sectors 12, 39, and 65, of which we can access only the first two at the time of writing. The PB binary exhibits prominent ETVs, suggesting dynamical interactions. The eclipse depths vary between the sectors due to systematics, and thus the true eclipse depths are larger than reported in Table \ref{tbl:main_table}. TIC 274862252 contaminates the lightcurve of the nearby, brighter TIC 274862251 at a separation of about 17 arcsec and magnitude difference of ${\rm \Delta T = -1.2 mag}$. In turn, TIC 274862252 is being contaminated by the nearby, fainter TIC 274862142, at a separation of about 44 arcsec and magnitude difference of $\rm {\Delta T = 0.47 mag}$. TIC 274862142 is itself an EB with P = 3.01 days -- very close to PB -- and is present in the lightcurve of TIC 274862252 as a weak signal.

\item[$\bullet$ TIC 285853156 (TGV-134)]: TESS observed the system in Sectors 43, 44, and 45, and detected two EBs with orbital periods of PA = 1.77 days and PB = 10 days, respectively. Both EBs exhibit prominent ETVs, which, in the case of PB are clearly anti-correlated -- and quite dramatic (see Fig. \ref{fig:285853156}) -- indicating strong dynamical interactions. Gaia measured AEN = 0.38 mas with an AESN of 182, and RUWE = 3.18, further strengthening the bound quadruple interpretation. Analysis of the system is in progress. We are currently collecting spectra from the Tillinghast Reflector Echelle Spectrograph (TRES; Szentgyorgyi \& Furesz 2007; Furesz 2008), which already show a sharp peak in the cross-correlation function corresponding to PB as well as hints of another peak -- weaker but broader -- likely due to the PA primary. 

\item[$\bullet$ TIC 370274336 (TGV-148)]: The target was observed by TESS in Sectors 18 and 58, and produced two sets of eclipses. One has a period of PA = 8 days. The other consists of two pairs of eclipses (one of which is blended with PA). One of the pairs is separated by 1089.87 days and the other by about 1089.85 days, indicating that the PB orbital period must be ${\rm \approx1089.86/N}$ days, where N is an integer (see Fig. \ref{fig:370274336}). Gaia measured AEN = 0.08 mas with an AESN of 9.9, and RUWE = 1.17, suggesting no significant orbital motion between the two EBs. 

\item[$\bullet$ TIC 37376063 (TGV-149)]: A quadruple candidate with PA = 3.96 days and PB = 14.53 days, observed by TESS in Sectors 8, 35, 61, and 62. Interestingly, the lightcurve from Sector 35 shows an extra pair of eclipses not associated with either EB (see Fig. \ref{fig:37376063}). Photocenter analysis confirms that these are on-target, suggesting that TIC 37376063 is in fact a (2+1)+2 quintuple system where the (2+1) triple hosts PA and is producing eclipses on its outer orbit as well. Analysis of the eclipse times shows prominent variations for the PA binary, as well as potential non-linearity for the PB binary. Gaia measured AEN = 0.9 mas, AESN = 423, and RUWE = 6.6, indicating potential orbital motion between the inner and outer components. Additionally, direct imaging observations show a resolved companion with a separation of $\sim0.1$ arcsec. Analysis of the system is in progress.

\item[$\bullet$ TIC 265274458 (TGV-41)]: A quadruple candidate listed in the K22 catalog, with PA = 2.99 days and PB = 57.33 days. The latter is based on two eclipses detected in Sectors 17 and 19 with similar depths and durations. Subsequent TESS observations show three additional eclipses in Sectors 52, 58, and 59 with different depths and durations. This contradicts the quadruple interpretation of K22, suggests that PB = 57.33 days is not an actual EB period and indicates that TIC 265274458 is a 2+1 triply-eclipsing triple star -- such that the tertiary produces outer eclipses and occultations on PA -- instead of a 2+2 quadruple system. We are pursuing follow-up observations of the target with TRES at the Fred L. Whipple Observatory. The current data suggests that the spectra might be double-lined.

\item[$\bullet$ TIC 282005870 (TGV-132)] TESS observed the target in Sectors 14, 41, 54, and 55. The two EBs have periods of PA = 0.68 days and PB = 1.79 days and produce both primary and secondary eclipses. Gaia measured AEN = 0.26 mas, AESN = 109, and RUWE = 1.8, and NSS = 1. More importantly, Gaia measured a complete orbital solution corresponding to an outer orbital period of 611 days, outer eccentricity of e = 0.557, and outer orbital inclination angle of 114 degrees, indicating that this is likely the outer orbit between PA and PB. 

\item[$\bullet$ TIC 167800999 (TGV-116)] The target was observed by TESS in Sectors 9, 36, and 63, producing two EBs with periods PA = 0.79 days and PB = 9.83 days. Gaia measured AEN = 1.53 mas, AESN = 2672, and RUWE = 10.97, as well as NSS = 3. The complete orbital solution from Gaia, likely between PA and PB, indicates an outer orbital period of 448 days with eccentricity of e = 0.156.

\item[$\bullet$ TIC 779824 (TGV-198)]: The target was observed by TESS in Sectors 8, 34, and 61, and produced two sets of eclipses (see Fig. \ref{fig:779824}). One follows a period of PA = 15.97 days and exhibits both primary and secondary eclipses. The other set consists of three eclipses observed in Sectors 34 and 61, two of which have similar depth ($\approx0.07$), duration ($\approx9.5$ hrs), and shape and the third is much shallower ($\approx0.02$). Overall, this suggests an EB producing two primary and one secondary eclipses, with a period PB ${\rm \approx749.29/N}$ days where N is an integer. Thus we include TIC 779824 in this catalog as a quadruple candidate, with the caveat that it could potentially be a triply-eclipsing triple like TIC 265274458. 

\item[$\bullet$ TIC 387288959 (TGV-156)]: TESS observed the target in 14 sectors and detected two sets of eclipses with PA = 2.7 days and PB = 83.1 days, both producing primary and secondary eclipses. Additionally, there is an extra pair of eclipses in Sector 59 separated by a few days and not associated with either PA or PB. The PA binary produces prominent ETVs with a turning point near the time of the extra pair of eclipses in Section 59, suggesting that PA is in fact a triply-eclipsing triple system (see Fig. \ref{fig:387288959}) and the extra pair of eclipses are in fact tertiary eclipses on the A binary. Gaia measured AEN = 0.75 mas, AESN = 518, and RUWE = 4.8, suggesting the detection of potential orbital motion. 

We note that there is nearby field star, TIC 1981408977, that is bright enough (${\rm \Delta T\approx0.65}$ mag) to produce either set of eclipses as contamination, yet too close for the photocenter analysis to distinguish between the two (1.4 arcsec, 0.07 pixels, Fig. \ref{fig:387288959}). The Gaia parallax and proper motions of TIC 387288959 and TIC 1981408977 are reasonably close (within the uncertainty) that the two may represent a bound system with a projected physical separation of about 700 AU (at a distance of about 660 pc). Multi-aperture principal component analysis of the FITSH lightcurve indicates that all eclipses, including the extra pair in Section 59, originate from TIC 387288959 but given the small separation between TIC 387288959 and TIC 1981408977 we cannot confirm the source. Thus the target is included in our catalog as the only exception for resolved systems due to it being either (i) a close (2+1)+2 quintuple on TIC 387288959 in a wide sextuple with TIC 1981408977 (or vice versa); or (ii) a close (2+1) triple on one of them in a wide quintuple with the PB EB on the other. Analysis of the system is in progress. We note that TIC 387288959 is a duplicate ID with TIC 1981408978. 

\item[$\bullet$ TIC 59453672 (TGV-177)]: The target was observed by TESS in Sectors 44 and 45, and produced two sets of eclipses with PA = 7.5 days and PB = 8.4 days, with the latter exhibiting potential ETVs. We have been obtaining spectroscopic observations with TRES and, to date, have collected 11 spectra. Some of them show as many as four distinct peaks in the cross-correlation function. 

\item[$\bullet$ TIC 284806955 (TGV-133)]: TESS observed the target in Sectors 17, 18, 24, and 58 and detected two sets of eclipses with PA = 1.89 days and PB = 2.46. As highlighted in Fig. \ref{fig:disentangle}, many of the eclipses are partially or fully-blended and the lightcurve had to be Fourier disentangled for accurate ephemerides measurements. Gaia measured AEN = 0.7 mas, AESN = 805, and RUWE = 4.4, indicating potential orbital motion between the two components. Additionally, the measured ETVs for both binaries show non-linear trends further suggesting dynamical interactions and a plausibly short outer orbital period.

\item[$\bullet$ Bright candidates with deep eclipses] To assist with the planning and scheduling of potential follow-up observations of interesting candidates by the community, we have provided in Table \ref{tab:good_for_obs} a list of targets brighter than T = 12 mag and primary eclipse depths greater than 1\%. Such observations can be aimed at e.g. long-term monitoring of eclipse times, constraining the presence of nearby resolved sources, radial velocity measurements, etc.

\end{description}

\section{Summary}
\label{sec:summary}

We have presented a catalog of eclipsing quadruple candidates discovered in TESS Full-Frame-Image data through a combination of machine-learning methods and visual inspection by citizen scientists. The catalog contains \Nquads targets producing two sets of eclipses following two distinct periods. All candidates have passed rigorous vetting tests, including pixel-by-pixel and photocenter analysis, and are considered to be close unresolved quadruple stars with a 2+2 hierarchical configuration. Some systems produce eclipse timing variations, in several cases quite dramatic, suggesting strong dynamical interactions between the two component eclipsing binary stars in a close orbit. Several systems show significant astrometric excess noise and high renormalized unit weight error as measured by Gaia, indicating the detection of astrometric motion in a wider quadruple. One target, TIC 37376063, exhibits all the above plus an additional pair of events associated with one of the EBs as tertiary eclipses, indicating a triply-eclipsing triple in a (2+1)+2 quintuple system. Another system, TIC 219006972, is the second most compact confirmed eclipsing quadruple star reported to date. Gaia measured non-single-star values of 1 for four targets, and an actual outer orbital period for one of them. We note that as Gaia measures more and longer orbits, a number of our quadruples candidates can be expected to have their outer orbits measured. The catalog provides a major increase in the number of thoroughly-vetted eclipsing quadruple candidates and highlights targets of interest that would benefit from further analysis. 

\clearpage
\acknowledgments
This paper includes data collected by the \emph{TESS} mission, which are publicly available from the Mikulski Archive for Space Telescopes (MAST). Funding for the \emph{TESS} mission is provided by NASA's Science Mission directorate.

Resources supporting this work were provided by the NASA High-End Computing (HEC) Program through the NASA Center for Climate Simulation (NCCS) at Goddard Space Flight Center.  Personnel directly supporting this effort were Mark L. Carroll, Laura E. Carriere, Ellen M. Salmon, Nicko D. Acks, Matthew J. Stroud, Bruce E. Pfaff, Lyn E. Gerner, Timothy M. Burch, and Savannah L. Strong.

This research has made use of the Exoplanet Follow-up Observation Program website, which is operated by the California Institute of Technology, under contract with the National Aeronautics and Space Administration under the Exoplanet Exploration Program. 

This research is based on observations made with the Galaxy Evolution Explorer, obtained from the MAST data archive at the Space Telescope Science Institute, which is operated by the Association of Universities for Research in Astronomy, Inc., under NASA contract NAS5-26555.

TB acknowledges the financial support of the Hungarian National Research, Development and Innovation Office -- NKFIH Grant KH-130372.

This work has made use of data from the European Space Agency (ESA) mission {\it Gaia} (\url{https://www.cosmos.esa.int/gaia}), processed by the {\it Gaia} Data Processing and Analysis Consortium (DPAC, \url{https://www.cosmos.esa.int/web/gaia/dpac/consortium}). Funding for the DPAC has been provided by national institutions, in particular the institutions participating in the {\it Gaia} Multilateral Agreement.

Resources supporting this work were provided by the NASA High-End Computing (HEC) Program through the NASA Advanced Supercomputing (NAS) Division at Ames Research Center for the production of the SPOC data products.

This work makes use of observations from the LCOGT network.

\facilities{
\emph{Gaia},
MAST,
TESS,
ASAS-SN,
ATLAS,
TRES,
ZTF
}

\software{
{\tt Astrocut} \citep{astrocut},
{\tt AstroImageJ} \citep{Collins:2017},
{\tt Astropy} \citep{astropy2013,astropy2018}, 
{\tt Eleanor} \citep{eleanor},
{\tt IPython} \citep{ipython},
{\tt Keras} \citep{keras},
{\tt LcTools} \citep{2019arXiv191008034S,2021arXiv210310285S},
{\tt Lightcurvefactory} \citep{2013MNRAS.428.1656B,2017MNRAS.467.2160R,2018MNRAS.478.5135B},
{\tt Lightkurve} \citep{lightkurve},
{\tt Matplotlib} \citep{matplotlib},
{\tt Mpi4py} \citep{mpi4py2008},
{\tt NumPy} \citep{numpy}, 
{\tt Pandas} \citep{pandas},
{\tt PHOEBE} \citep{2011ascl.soft06002P},
{\tt Scikit-learn} \citep{scikit-learn},
{\tt SciPy} \citep{scipy},
{\tt Tensorflow} \citep{tensorflow},
{\tt Tess-point} \citep{tess-point}
{\tt wotan} \citep{2019AJ....158..143H}
}

\clearpage

\section{Data Availability}

The data underlying this article will be shared on reasonable request to the corresponding author.

\bibliography{refs}{}
\bibliographystyle{aasjournal}

\newpage

\begin{figure}[h]
    \centering
    \includegraphics[width=0.995\linewidth]{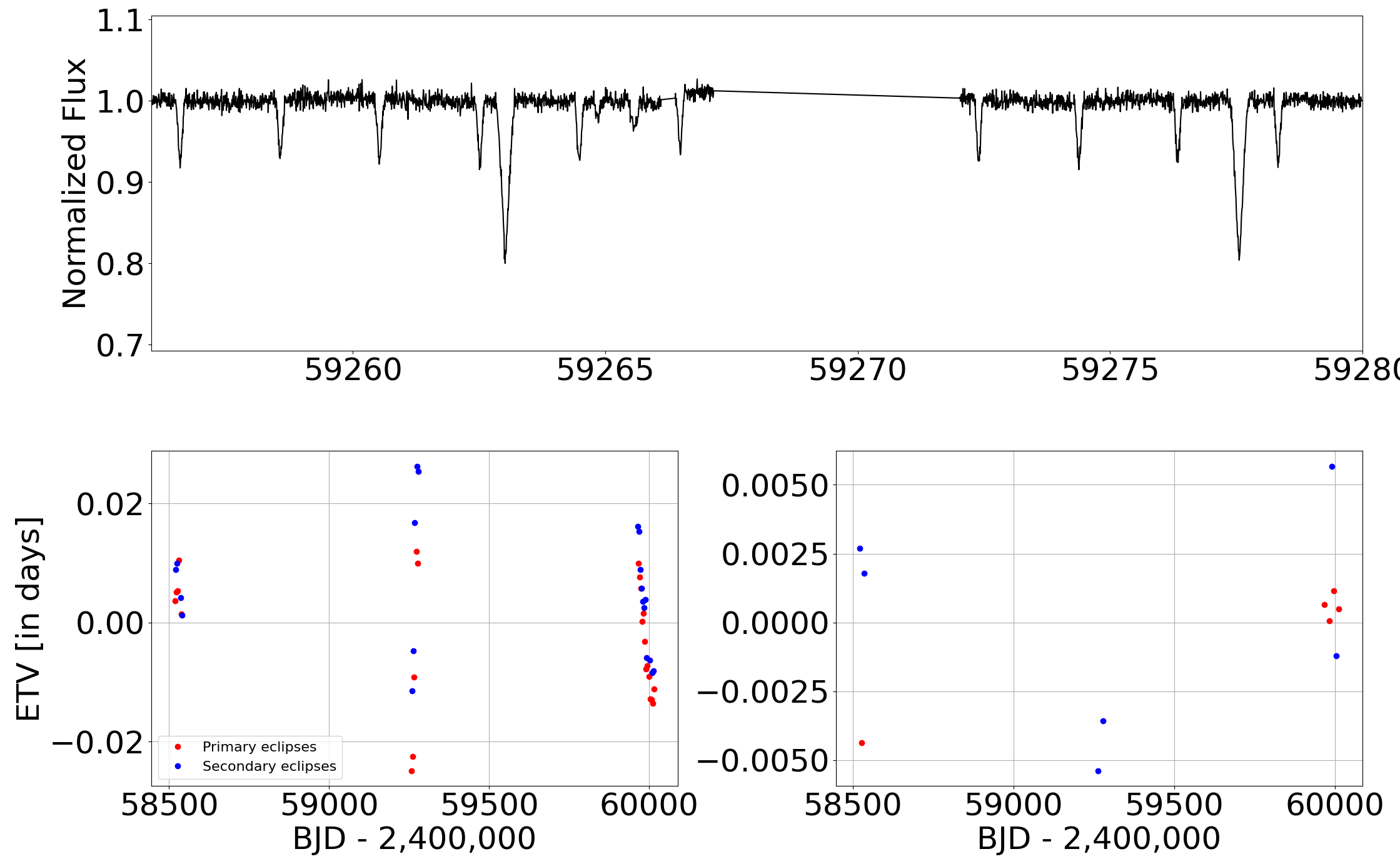}
    \caption{Upper panel: TESS lightcurve of TIC 387288959, highlighting the PA eclipses (red), one primary and one secondary of PB (blue) and two extra tertiary events (one on each side of the dashed cyan line). Lower panels: Measured eclipse timing variations for TIC 387288959 for the PA (left) and PB (right) components, in terms of Observed minus Calculated time as a function of the Observed time. The left panel shows a likely outer period of about 1300 days for the (2+1) triple containing PA = 2.7 days. The vertical dashed line in the lower left panel indicates the time around the two tertiary events.}
    \label{fig:387288959}
\end{figure}

\begin{figure}
    \centering
    \includegraphics[width=0.8\linewidth]{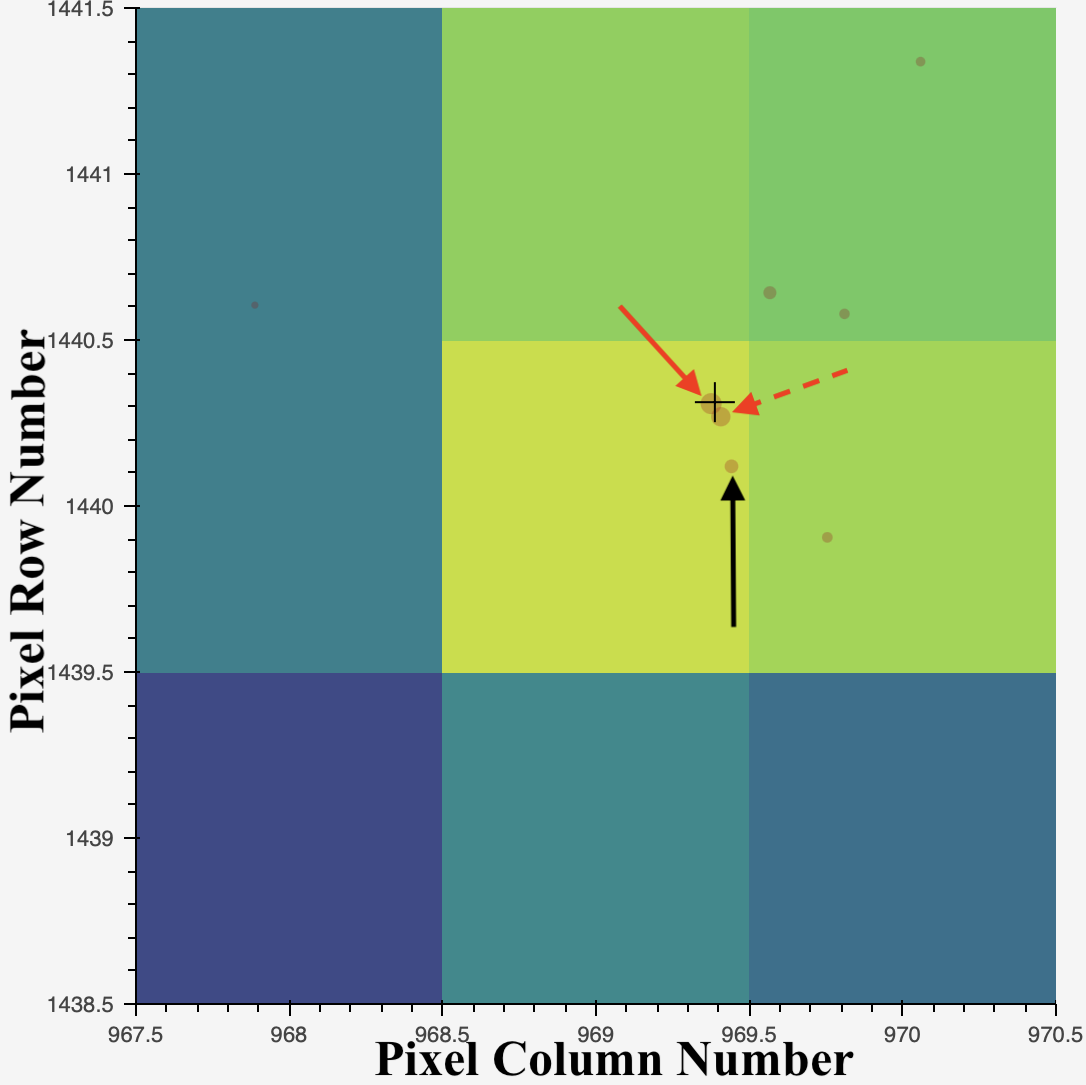}
    \includegraphics[width=0.48\linewidth]{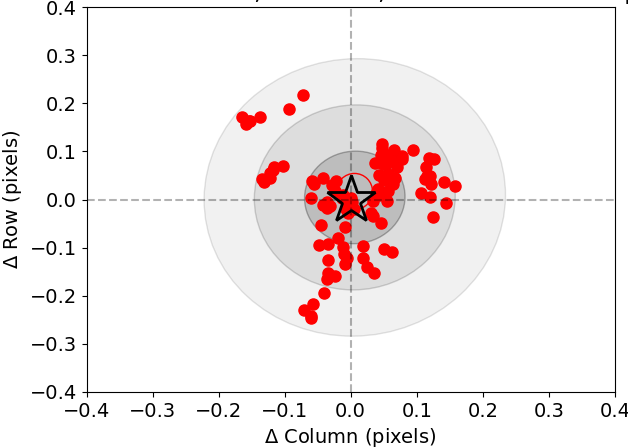}
    \includegraphics[width=0.48\linewidth]{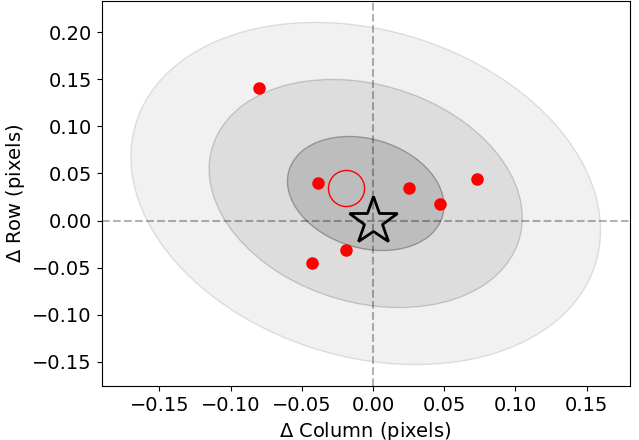}
    \caption{Upper panel: $3 \times 3$ TESS pixels Skyview image of TIC 387288959 (red solid arrow) highlighting nearby field stars. The red dashed arrow indicates TIC 1981408977, separated by 1.4 arcsec (0.07 pixels) and having a magnitude difference of ${\rm \Delta T\approx0.65}$ mag; the black solid arrow indicates TIC 1981408976, with a separation of 4.4 arcsec (0.2 pixels) and ${\rm \Delta T\approx3}$ mag. Both of them are potential sources of contamination. Lower panels: Photocenter analysis for TIC 387288959. The small filled symbols represent the individual measured photocenters for each unblended eclipse of PA = 2.7 days (left panel) and PB = 83.1 days (right panel) for all sectors. The large open circles represent the measured average photocenter and the grey contours represent the corresponding 1-, 2- and $3-\sigma$ confidence intervals. While the photocenter analysis firmly rules out TIC 1981408976 (black arrow) as a potential source of the detected eclipses, TIC 387288959 (red solid arrow) and TIC 1981408977 (red dashed arrow) are too close to each other for reliable source determination.}
    \label{fig:387288959_photocenter}
\end{figure}

\newpage
\clearpage

{\onecolumngrid
\begin{longtable}{llllllllrrrr}

\caption{Parameters of the \Nquads quadruple candidates presented here, including ephemerides, eclipse depths and durations, and additional comments. The full table is available online.}\\

\hline
\hline
TIC ID & RA & Dec & Binary & Period & T${_0}$ & Phase$_{s}$ & Dep$_{p}$ & Dep$_{s}$ & Dur$_{p}$ & Dur$_{s}$ \\
- & degrees & degrees & - & d & BJD-2457000 & & ppt & ppt & hr & hr\\
\hline
\endhead

274862252 & 237.426543 & -53.506278 & A & 2.3437 & 1627.085 & $0.505\pm0.001$ & 9 & 4 & 5.0 & 4.0\\
& & & B & 2.9531 & 2363.6153 & $0.3617\pm0.0004$ & 17 & 13 & 5.0 & 6.0\\
\multicolumn{11}{l}{Additional information: TGV-98, Gaia DR3 5885035386669240960, Tmag: 13.50, Teff: 7287 K, Dist: 4324.52 pc}\\
\multicolumn{11}{l}{Comments: Contaminator for TIC 274862251; heavily-blended eclipses; prominent ETVs on PB; heavy contamination}\\
\multicolumn{11}{l}{~~~~~~~~~~~~~~~~from TIC 274862142 (nearby EB with P = 3.01 days); ephemerides might be slightly off;}\\
\multicolumn{11}{l}{~~~~~~~~~~~~~~~~true depths are much larger; depth differences between sectors}\\
\multicolumn{11}{l}{ Sectors observed: 12, 39, 65}\\
\hline
120911334 & 73.862103 & 38.312956 & A & 2.0005 & 1820.0645 & $0.5963\pm0.0002$ & 124 & 77 & 3.6 & 4.0\\
& & & B & 2.4536 & 1819.2211 & $0.5029\pm0.0003$ & 139 & 115 & 3.2 & 4.5\\
\multicolumn{11}{l}{Additional information: TGV-99, Gaia DR3 199017518802605440, Tmag: 14.59, Teff: --, Dist: --}\\
\multicolumn{11}{l}{Comments: --}\\
\multicolumn{11}{l}{ Sectors observed: 19, 59}\\
\hline
121511673 & 74.971948 & 40.402072 & A & 2.1590 & 1819.8709 & $0.4955\pm0.0013$ & 126 & 85 & 5.3 & 5.0\\
& & & B & 3.7872 & 1820.2514 & $0.4982\pm0.0007$ & 113 & 81 & 6.1 & 5.6\\
\multicolumn{11}{l}{Additional information: TGV-100, Gaia DR3 200791129837694848, Tmag: 13.76, Teff: 11891 K, Dist: 3028.07 pc}\\
\multicolumn{11}{l}{Comments: blended eclipses;}\\
\multicolumn{11}{l}{ Sectors observed: 19, 59}\\
\hline
125787704 & 115.159669 & -27.319573 & A & 6.5593 & 1500.2603 & $0.5028\pm0.0004$ & 33 & 15 & 5.4 & 5.4\\
& & & B & 7.6917 & 1497.7007 & $0.4662\pm0.0003$ & 44 & 19 & 5.5 & 5.1\\
\multicolumn{11}{l}{Additional information: TGV-101, Gaia DR3 5612234697703939200, Tmag: 13.06, Teff: 7968 K, Dist: 2429.22 pc}\\
\multicolumn{11}{l}{Comments: depth differences between sectors; blended eclipses;}\\
\multicolumn{11}{l}{ Sectors observed: 7, 34, 61}\\
\hline
127011022 & 116.506091 & -25.471756 & A & 2.2689 & 1494.3224 & $0.4983\pm0.0012$ & 12 & 11 & 2.5 & 2.4\\
& & & B & 2.8549 & 1506.1814 & $0.5013\pm0.0017$ & 18 & 15 & 3.1 & 3.6\\
\multicolumn{11}{l}{Additional information: TGV-102, Gaia DR3 5614373041669940480, Tmag: 13.31, Teff: --, Dist: 1552.26 pc}\\
\multicolumn{11}{l}{Comments: \textsc{eleanor} aperture is off-target; depth differences between sectors; depths quoted for Sector 34; potential ETVs;}\\
\multicolumn{11}{l}{~~~~~~~~~~~~~~~~PB potentially half}\\
\multicolumn{11}{l}{ Sectors observed: 7, 34, 61}\\
\hline
128802666 & 118.418216 & -27.662181 & A & 1.6932 & 1496.8473 & -- & 17 & -- & 3.8 & --\\
& & & B & 4.8618 & 1492.6553 & $0.7776\pm0.0010$ & 55 & 45 & 5.8 & 5.5\\
\multicolumn{11}{l}{Additional information: TGV-103, Gaia DR3 5600765069135007616, Tmag: 13.65, Teff: 10916 K, Dist: 6896.94 pc}\\
\multicolumn{11}{l}{Comments: Crowded field; depth differences between sectors; potential ETVs on B}\\
\multicolumn{11}{l}{ Sectors observed: 7, 8, 34, 61}\\
\hline
13021681 & 75.534801 & -24.061641 & A & 0.5406 & 2175.6784 & -- & 40 & -- & 1.0 & --\\
& & & B & 11.0651 & 1445.7904 & $0.2435\pm0.0007$ & 360 & 170 & 3.1 & 3.7\\
\multicolumn{11}{l}{Additional information: TGV-104, Gaia DR3 2960631977745576064, Tmag: 13.93, Teff: 4173 K, Dist: 361.04 pc}\\
\multicolumn{11}{l}{Comments: depth differences between sectors; A secondary too low SNR for reliable measurements}\\
\multicolumn{11}{l}{ Sectors observed: 5, 32}\\
\hline

\multicolumn{11}{c}{{------------------------------------------------------------------}}\\
\multicolumn{11}{c}{{Continued on next page}}\\
\multicolumn{11}{c}{{TGV-N = TESS/Goddard/VSG quadruple candidate -N, Phase$_{s}$ = Secondary phase}}\\
\multicolumn{11}{c}{{Dep$_{n}$ = Depth of eclipse $n$, Dur$_{n}$ = Duration of eclipse $n$}}\\
\multicolumn{11}{c}{{Teff = Composite effective temperature, ppt = parts-per-thousand}}\\


\label{tbl:main_table}
\end{longtable}
}

\clearpage


\begin{table}
\begin{tabular}[t]{cc}   
\begin{tabular}[t]{c|c|c|c} 
\hline
\hline
TIC & Tmag & ${\rm Depth_A (\%)}$ & ${\rm Depth_B (\%)}$ \\
\hline
153406662 & 8.04 & 2 & 0.5 \\
444816203 & 8.89 & 2 & 8 \\
238558210 & 9.12 & 3 & 3 \\
167800999 & 9.24 & 4 & 28 \\
320233974 & 9.52 & 3 & 5 \\
278465736 & 9.86 & 9 & 13 \\
346000664 & 10.04 & 12 & 2 \\
125583594 & 10.10 & 1 & 3 \\
356318101 & 10.15 & 3 & 10 \\
329089161 & 10.27 & 0.5 & 2 \\
459333241 & 10.70 & 1 & 2 \\
321474625 & 10.76 & 3 & 4 \\
429901945 & 10.80 & 28 & 14 \\
407060024 & 10.94 & 1 & 11 \\
146435300 & 10.96 & 1 & 1 \\
140328928 & 11.02 & 4 & 1 \\
284806955 & 11.11 & 3 & 1 \\
270360534 & 11.24 & 8 & 7 \\
282005870 & 11.30 & 4 & 7 \\
285853156 & 11.47 & 5 & 21 \\
405114468 & 11.56 & 3 & 12 \\
59453672 & 11.60 & 1 & 3 \\
304713857 & 11.69 & 13 & 10 \\
80893927 & 11.91 & 1 & 1 \\
47279366 & 11.97 & 3 & 21 \\
160514618 & 11.97 & 11 & 15 \\
24700485 & 11.99 & 0.5 & 7 \\
\hline
\end{tabular} &  
\end{tabular}

\caption{Quadruple candidates brighter than Tmag = 12 and primary eclipse depths greater than 1\% for one of both EBs (sorted by magnitude).}
\label{tab:good_for_obs}
\end{table}

\appendix
\setcounter{figure}{0}
\renewcommand{\thefigure}{A\arabic{figure}}

\section{Figures}

\begin{figure}[h]
    \centering
    \includegraphics[width=0.99\linewidth]{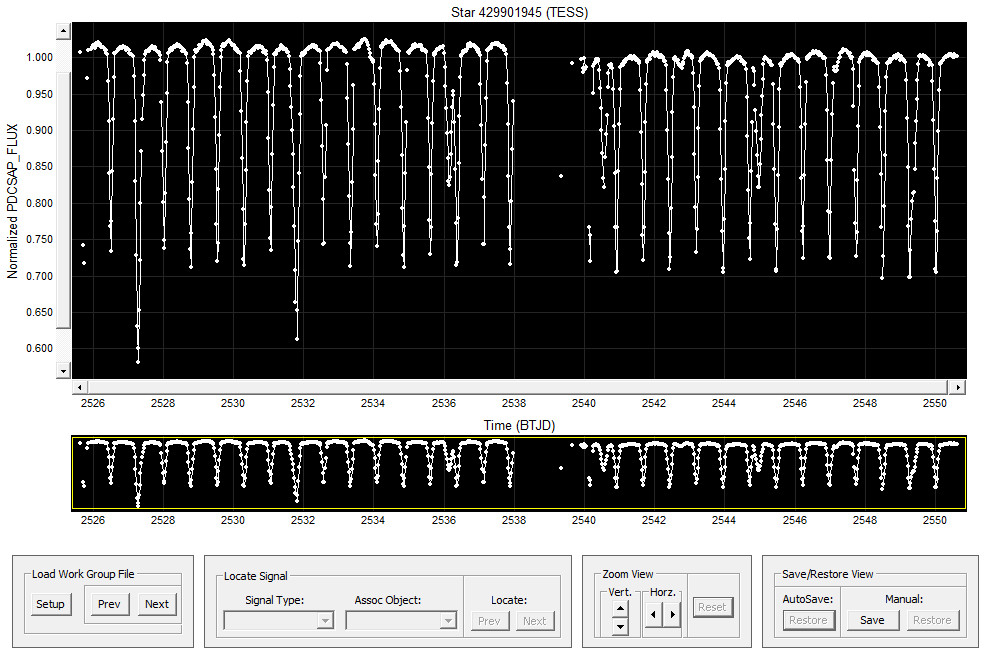}
    \caption{Example screenshot from LcViewer showing the Sector 45 TESS Full-Frame Image lightcurve used to discover the quadruple candidate TIC 429901945.}
    \label{fig:LcTools}
\end{figure}

\begin{figure}
    \centering
    \includegraphics[width=0.99\linewidth]{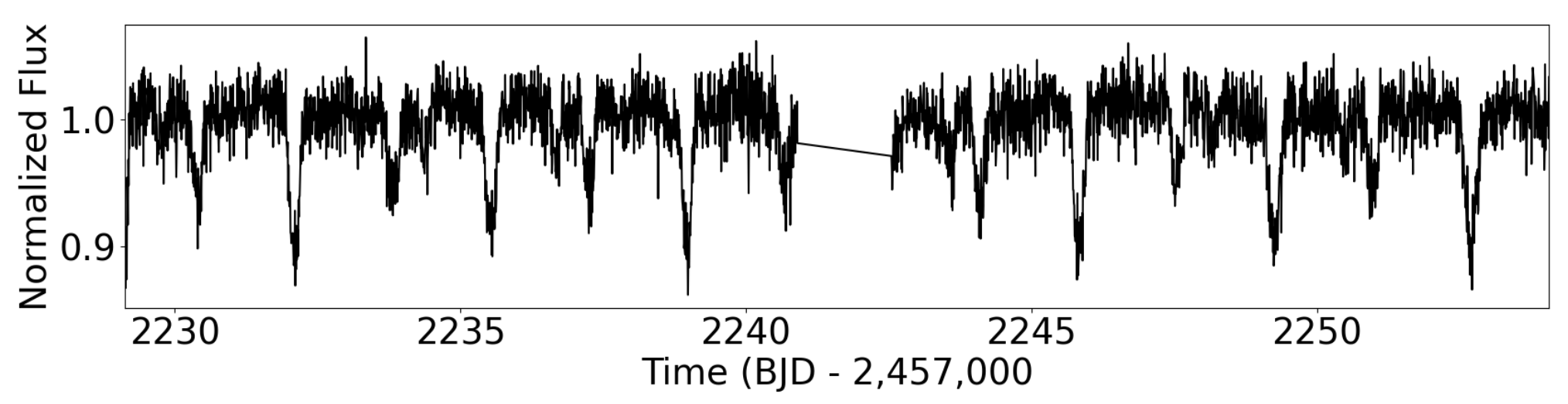}
    \includegraphics[width=0.95\linewidth]{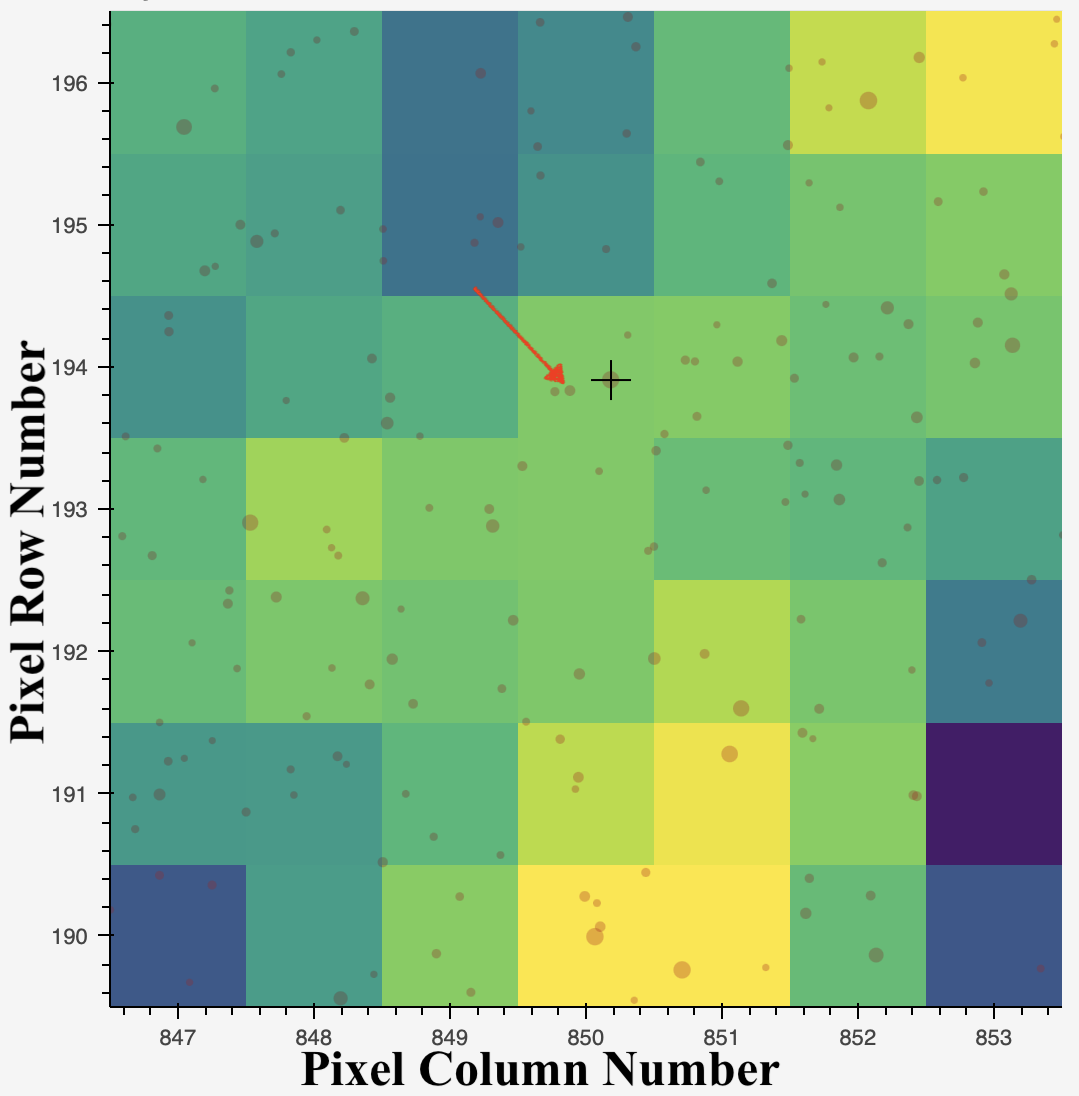}
    \caption{Upper panel: TESS \textsc{eleanor} lightcurve for TIC 144475902 (G = 14.5 mag) from Sector 34. Lower panel: Corresponding $7 \times 7$ TESS pixels Skyview image of the field around the target (black cross symbol) showing resolved nearby stars brighter than G = 21 mag. The red arrow indicates a field star that is bright enough be a source of contamination (TIC 144475903), yet too close to the target to be ruled out from the pixel-by-pixel analysis. As discussed below, sub-pixel photocenter analysis rules out TIC 144475903 and confirms that TIC 144475902 is the source of both EBs.}
    \label{fig:pixel_by_pixel_1}
\end{figure}

\begin{figure}
    \centering
        \includegraphics[width=0.99\linewidth]{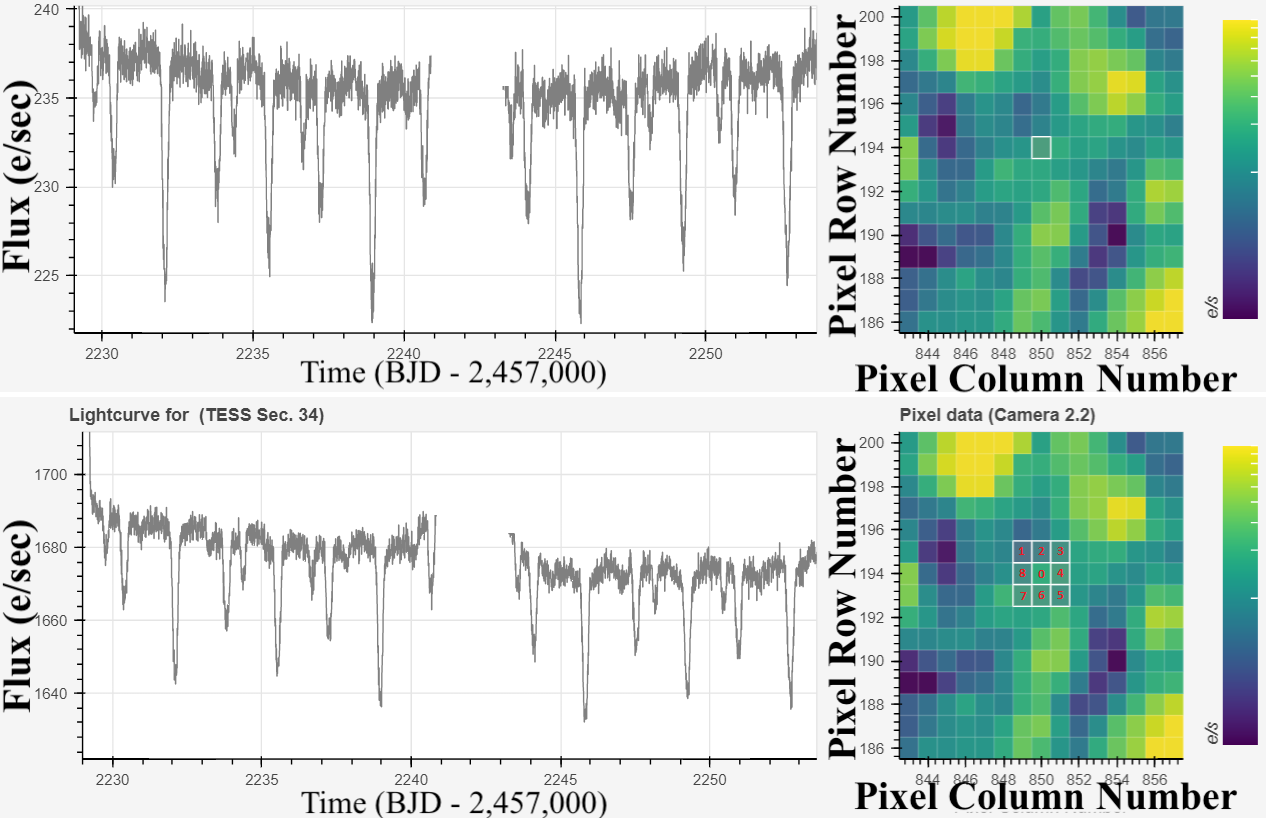}
    \caption{Lightkurve pixel-by-pixel analysis for the quadruple candidate TIC 144475902. The right panels represent the TESS field of view, highlighting the pixels (white contours) used to extract the lightcurve shown in the left panel. Both sets of eclipses scale equally in the surrounding pixels (labelled 1 through 8 in the lower right panel), indicating that the target is their source. }
    \label{fig:pixel_by_pixel_2}
\end{figure}

\begin{figure}
    \centering
    \includegraphics[width=0.49\linewidth]{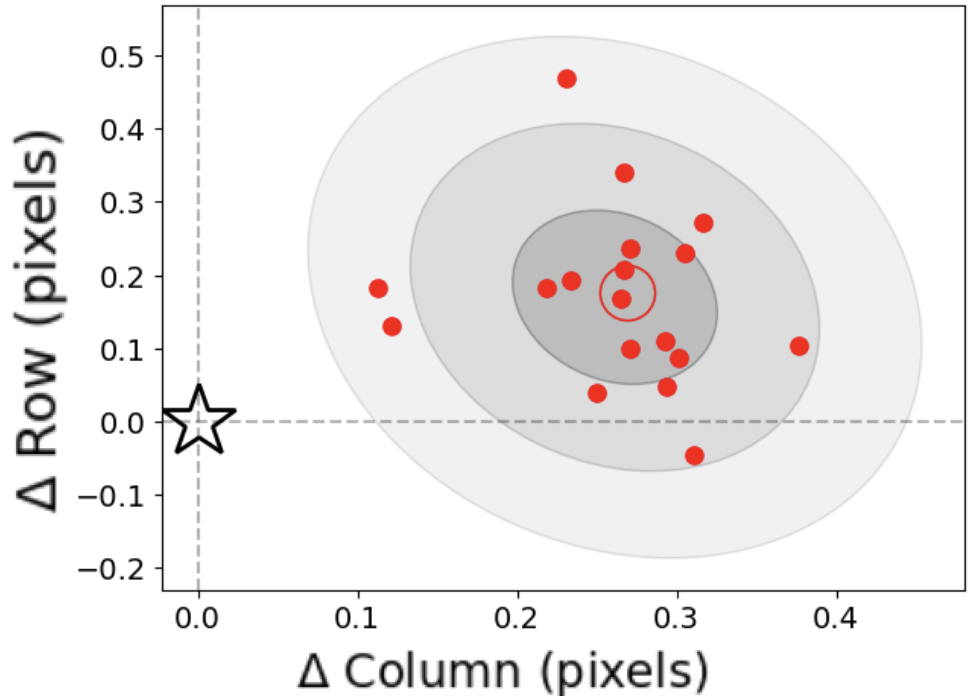}
    \includegraphics[width=0.49\linewidth]{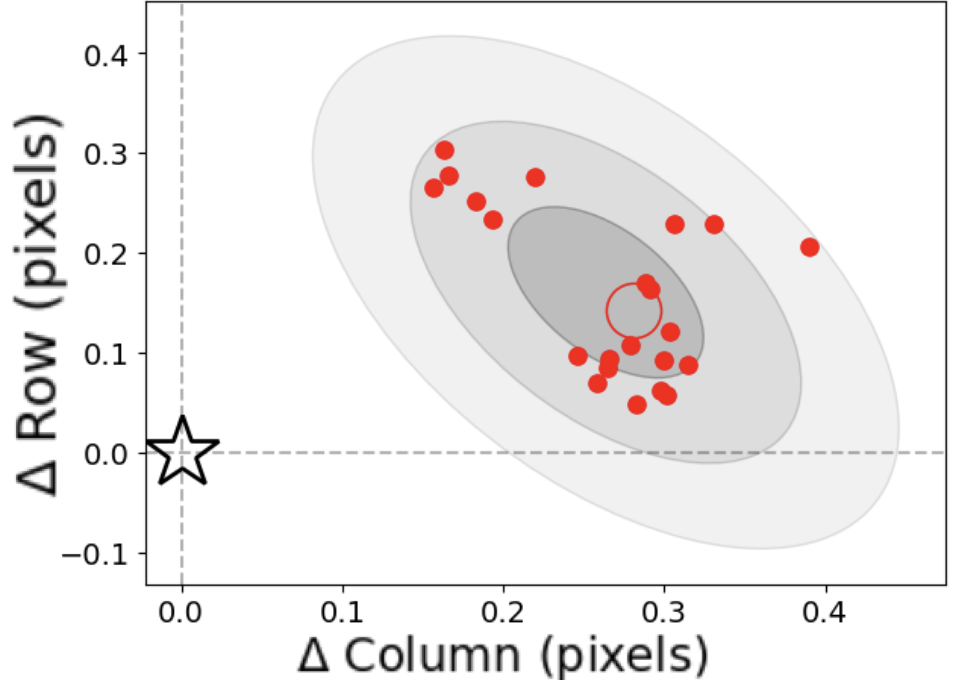}
    \caption{Photocenter analysis for TIC 144475903, a potential source of contamination for the quadruple candidate TIC 144475902. The small filed symbols represent the individual measured photocenters for each unblended eclipse of PA = 2.3 days (left panel) and PB = 3.43 days (right panel) for all sectors (7, 8, 34, 61). The large open circles represent the measured average photocenter and the grey contours represent the corresponding 1-, 2- and $3-\sigma$ confidence intervals. The measured photocenters for both EBs move away from the position of TIC 144475903 (black star symbol) and towards the position of TIC 144475902 (located near the large open symbol, about 0.3 pixels away from TIC 144475903), confirming that TIC 144475902 is the source of PA and PB.}
    \label{fig:144475903}
\end{figure}

\begin{figure}
    \centering
    \includegraphics[width=0.95\linewidth]{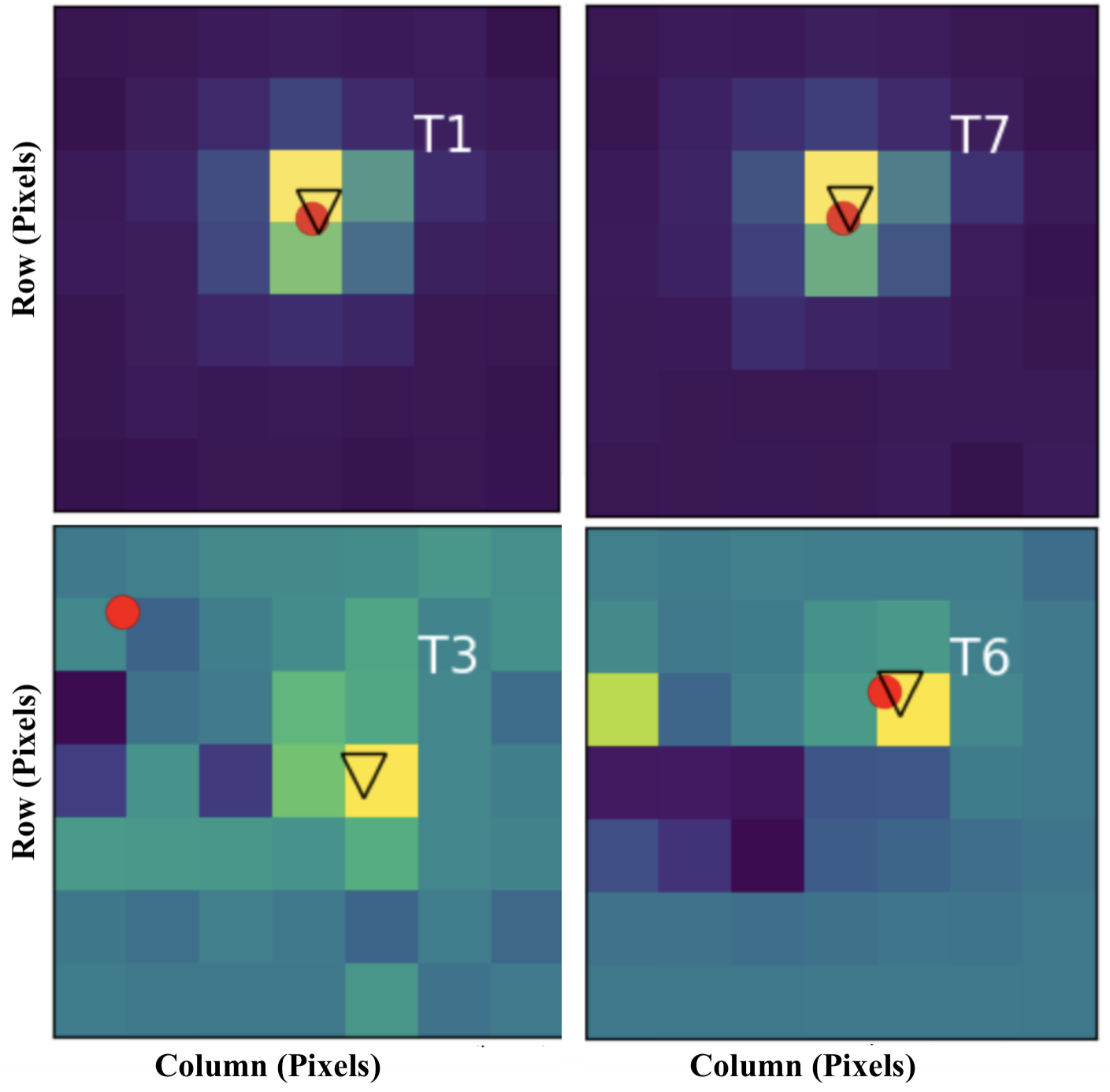}
    \caption{Example of difference images. Upper panels: Good-quality difference images (for TIC 25712085, eclipses 1 and 7, sector 9) showing a single bright spot superimposed on a dark background. The upside-down triangles represents the PSF-measured photocenter, and the red dot represents the Pixel-Response-Function-measured photocenter. Lower panels: Difference images dominated by systematics (TIC 256197811, eclipses 3 and 6, sector 24). The lower left panel shows a successful PSF fit and unsuccessful PRF whereas both fits are reported successful for the difference image shown in the lower right panel. However, it is clear that in both cases the difference images are poor, the respective photocenter measurements are unreliable, and have to be removed from consideration for the photocenter analysis.}
    \label{fig:bad_diff}
\end{figure}

\begin{figure}
    \centering
    \includegraphics[width=0.95\linewidth]{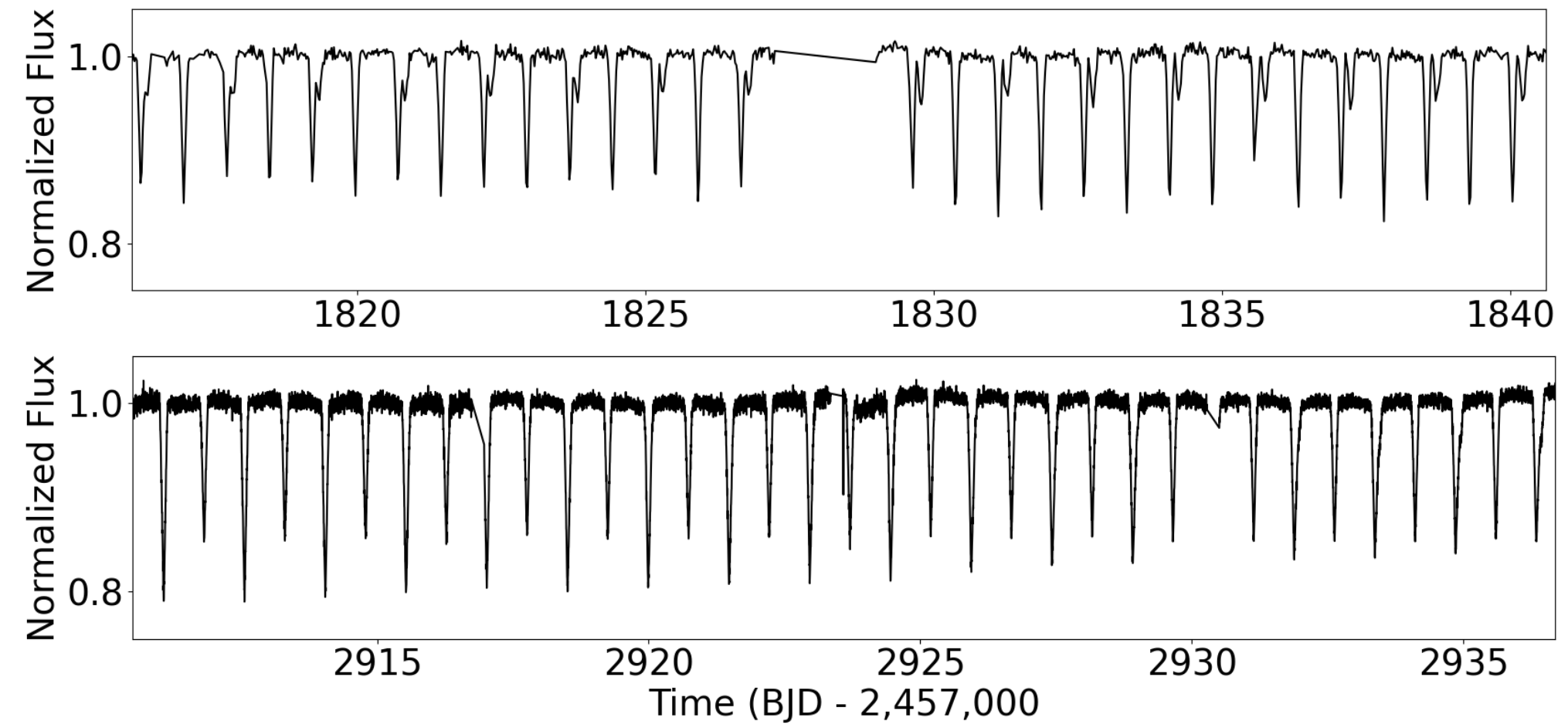}
    \includegraphics[width=0.75\linewidth]{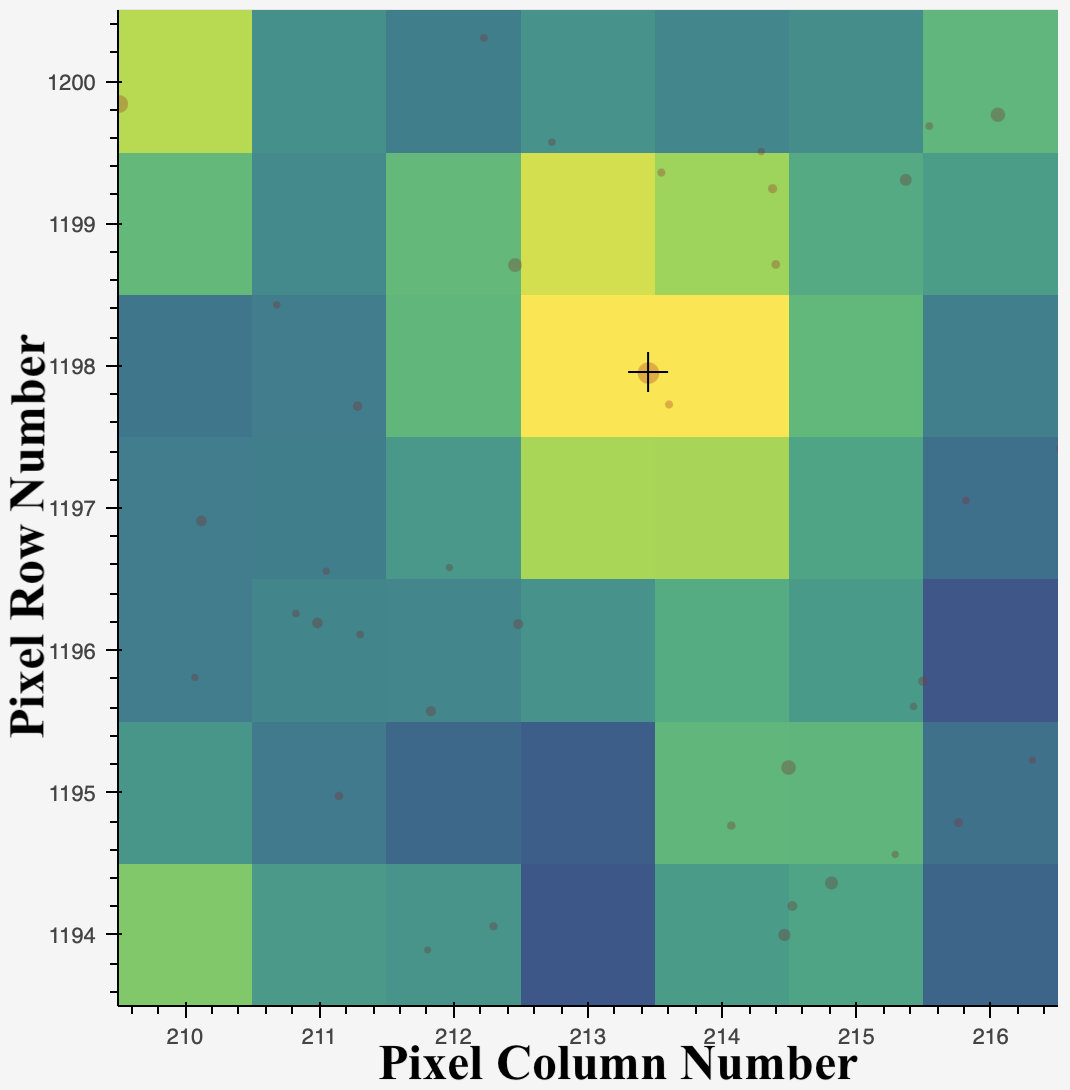}
    \caption{Quadruple candidate TIC 252301701 exhibiting two sets of eclipses with nearly-identical periods $PA\approx1.487$ days and $PB\approx1.491$ days. The upper and middle panels show the \textsc{eleanor} lightcurve for Sector 9 (upper) and Sector 59 (lower), respectively. Lower panel: $7 \times 7$ TESS pixels Skyview image centered on the target. The overlap between the two sets of eclipses is partial in Sector 9. However, the nearly-complete blend between the two EBs in Sector 59, can be exploited to our benefit since the lightcure practically looks like that of a single EB. Photocenter analysis confirms that the Sector 59 eclipses are on-target, indicating that this is indeed a genuine quadruple candidate. With that said, TIC 252301701 was independently discovered by \cite{2022MNRAS.517.2190R} and is thus excluded from our catalog.}
    \label{fig:VDLC}
\end{figure}

\newpage
\clearpage

\end{document}